\documentclass[twocolumn,tighten]{aastex701}
\input{preamble.sty}
\begin{document}
\title{Radiative relativistic reconnection powering the hard-state emission of X-ray binaries: \\synchrotron-driven pair loading, MeV tail, and seed photons}

\correspondingauthor{Valeriia~Rohoza}
\email{valeriia.rohoza@columbia.edu}

\author[0009-0000-4703-9808]{Valeriia~Rohoza}
\affiliation{Department of Astronomy and Columbia Astrophysics Laboratory, Columbia University, New York, NY 10027, USA}
\email{valeriia.rohoza@columbia.edu}

\author[0000-0001-6640-0179]{Maria~Petropoulou}
\affiliation{Department of Physics, National and Kapodistrian University of Athens, University Campus Zografos, GR 15784, Athens, Greece}
\email{mpetropo@phys.uoa.gr}

\author[0000-0001-8939-6862]{Hayk~Hakobyan}
\affiliation{Center for Computational Astrophysics, Flatiron Institute, 162 5th avenue, New York, NY 10010, USA}
\affiliation{Department of Astronomy and Columbia Astrophysics Laboratory, Columbia University, New York, NY 10027, USA}
\email{haykh.astro@gmail.com}

\author[0000-0002-2383-3860]{Despina~Karavola}
\affiliation{Department of Physics, National and Kapodistrian University of Athens, University Campus Zografos, GR 15784, Athens, Greece}
\email{dkaravola@phys.uoa.gr}

\author[0000-0002-1227-2754]{Lorenzo~Sironi}
\affiliation{Department of Astronomy and Columbia Astrophysics Laboratory, Columbia University, New York, NY 10027, USA}
\affiliation{Center for Computational Astrophysics, Flatiron Institute, 162 5th avenue, New York, NY 10010, USA}
\email{lsironi@astro.columbia.edu}

\begin{abstract}
In the hard state, X-ray emission from black hole X-ray binaries is attributed to Comptonization in the corona, a highly magnetized region near the black hole. Relativistic magnetic reconnection drives bulk motions that can upscatter soft photons and reproduce the spectral peak at $\sim 100$~keV for parameters relevant to Cygnus~X-1. However, the origin of the seed photons and of the MeV tail above the peak remains unclear. In 3D reconnection, so-called ``free'' particles escape the flux ropes and accelerate at the maximum rate set by the reconnection electric field. Their synchrotron emission extends to tens of MeV, which enables $\gamma \gamma$ pair production with the Compton-upscattered photons. We present 3D particle-in-cell simulations of relativistic reconnection in pair plasma, in which particles cool by inverse Compton (IC) scattering below the synchrotron self-absorption threshold and by synchrotron radiation above it. Of the incoming magnetic power, $\sim 40 \%$ is converted to IC radiation and $\gtrsim 10$--$20\%$ to synchrotron. Using analytical estimates and radiative calculations for Cygnus~X-1, we model how the radiation from reconnection-accelerated particles is reprocessed in the corona. Pair production attenuates the synchrotron emission above $\sim 1$~MeV, and the escaping $\gamma$-rays form an MeV tail consistent with the observations, extending up to at least $\sim 10$~MeV. Most of the absorbed power goes into secondary pairs, whose Thomson optical depth can reach order unity. Their synchrotron emission, peaking at $\sim 0.1 \, \text{keV}$, can supply the seed photons for Comptonization. Our results support reconnection-based synchrotron self-Compton models of the hard-state emission in X-ray binaries.
\end{abstract}
\section{Introduction}\label{sec:intro}
\setcounter{footnote}{0}
Black hole (BH) X-ray binaries, such as Cygnus~X-1, exhibit two main spectral states: soft and hard (see \citealt{remillard_xray_2006} for a review). The soft-state continuum is explained by thermal emission from the accretion disk. The hard state is thought to be powered by the BH corona; its spectral energy distribution $\varepsilon L_\varepsilon$ peaks at $\varepsilon \sim 100 \, \text{keV}$ and extends into an MeV tail \citep{mcconnell_soft_2002,2017A&A...603A...8W,2017MNRAS.471.3657Z}. \citet{2017MNRAS.471.3657Z} suggest that emission detected up to $\varepsilon \sim 0.1 \, \text{GeV}$ may share the same origin.

Inverse Compton (IC) scattering of low-energy ``seed'' photons by hot coronal electrons with Thomson optical depth $\tau_T \sim 1$ naturally explains the spectral peak at $\sim 100 \, \text{keV}$ \citep{mcconnell_soft_2002}. The MeV tail, however, requires an additional, likely non-thermal, particle population (\citealt{1998PhST...77...57P}; see also Section~3.1.1 of \citealt{1998tbha.conf..100P}). Models containing both thermal and non-thermal particles leave three key questions unanswered: What heats the coronal electrons to $kT \sim 100 \, \text{keV}$ and sustains this temperature against radiative losses? What produces the non-thermal particles responsible for the MeV tail? What is the origin of the seed photons?

The seed photons may stem from the accretion disk or from synchrotron emission within the corona; \citet{2009ApJ...690L..97P} advocate the latter. Because synchrotron photons are produced locally, they are more likely to interact with the Comptonizing plasma than disk photons \citep{2014SSRv..183...61P}. Moreover, disk photons with energies above $\varepsilon \sim 1 \, \text{keV}$ would not be absorbed by the interstellar medium and should appear as an additional spectral component; in the hard state of some sources, this component is weak or absent \citep[e.g.,][]{2001ApJ...555..483E}. For the MeV tail, \citet{2009ApJ...690L..97P} invoke Compton scattering of synchrotron photons by non-thermal electrons, but do not specify the electron acceleration mechanism.

Bulk motions driven by magnetic reconnection or turbulence offer an alternative to the thermal-plasma picture. Both processes convert electromagnetic energy into plasma energy and generate bulk motions capable of Comptonizing seed photons (reconnection: e.g., \citealt{beloborodov_radiative_2017,sironi_kinetic_2020,sridhar_comptonization_2021,sridhar_comptonization_2023,gupta_comptonization_2024}; turbulence: \citealt{groselj_radiative_2024,nattila_radiative_2024,2026ApJ..1001...64G}). Motivated by turbulence simulations, \citet{2026arXiv261009676L} recently reproduced the $\sim 100 \, \text{keV}$ peak and the MeV tail using thermal and non-thermal electron populations, respectively, with accretion-disk photons as seeds. For their adopted parameters, we estimate that synchrotron emission extends only to $\sim 0.2 \, \text{MeV}$. Thus, as in \citet{2009ApJ...690L..97P}, the MeV tail in their model is produced by IC scattering off non-thermal electrons.

Distinguishing observationally between the reconnection and turbulence scenarios remains challenging. In turbulence, bulk motions depend sensitively on the strength of the fluctuating fields relative to the mean field \citep{2018PhRvL.121y5101C}. By contrast, in relativistic reconnection \citep[see][for a review]{sironi_relativistic_2025}, plasmoid bulk motions are remarkably insensitive to plasma magnetization, provided that the upstream magnetic enthalpy exceeds the relativistic plasma enthalpy \citep{sridhar_comptonization_2021}. This robustness---perhaps desirable given the apparent universality of coronal conditions across systems---motivates us to focus on relativistic reconnection and adopt the plasmoid-chain Comptonization model of \citet{beloborodov_radiative_2017} for the hard-state emission.

In this model, particles are rapidly accelerated at reconnection sites (X-points) and become trapped in plasmoids, where they cool. Using reconnection simulations with seed photons injected in the plane of the current sheet, \citet{sridhar_comptonization_2021} showed that this model can reproduce the Comptonized emission up to the $\sim 100\,\text{keV}$ peak. Motivated by 2D simulations of relativistic reconnection, the model assumes that no fast particle acceleration occurs beyond the X-point phase \citep{beloborodov_radiative_2017}. As a result, most of the observed emission is  produced by particles with energies below the synchrotron self-absorption (SSA) threshold, such that synchrotron losses are negligible.

Recent 3D simulations of relativistic reconnection reveal a novel acceleration mechanism that enables particles to reach energies far exceeding those found in 2D simulations \citep{zhang_fast_2021,zhang_origin_2023,chernoglazov_high_2023}. In 3D, flux ropes have a finite extent along the reconnection electric field, allowing particles to escape and be linearly accelerated---at nearly the maximum rate set by the reconnection electric field---while meandering between the two upstream sides of the layer \citep{zhang_fast_2021,zhang_origin_2023}. 
These ``free'' particles may eventually be recaptured by the reconnection downstream, but during their ``free'' phase they can reach energies far above the SSA threshold, making synchrotron emission their dominant cooling channel. In fact, they can be accelerated up to the so-called synchrotron burnoff limit and radiate photons with energies up to tens of MeV. This physics was absent from the model of \citet{beloborodov_radiative_2017}, which was motivated by 2D simulations. 

Therefore, the reconnection-based model for the hard state of X-ray binaries needs to be revisited. In this work, we address the following questions: Can synchrotron emission from rapidly accelerated particles in 3D reconnection shape the MeV tail of the emerging spectrum? Can the resulting secondary pairs produce a Thomson optical depth of order unity? Can synchrotron emission from these pairs provide a competitive source of seed photons for Comptonization?

Using Cygnus~X-1 as an example, we quantify the role of high-energy synchrotron photons in the coronal emission of X-ray binaries, assuming a pair-plasma composition. In Section~\ref{sec:rad-rec}, we establish the characteristic energy scales and show that synchrotron photons can reach the burnoff limit of $\sim 16$~MeV and therefore pair-produce with IC photons. In Section~\ref{sec:analytical}, we analytically estimate the products of $\gamma\gamma$ interactions. In Section~\ref{sec:pic}, we quantify how reconnection partitions the released magnetic energy between IC and synchrotron emission. In Section~\ref{sec:lehamoc}, we numerically model $\gamma\gamma$ pair production in the corona. We discuss our results and next steps in Section~\ref{sec:disc} and conclude in Section~\ref{sec:sum}.

\section{Radiative relativistic reconnection}\label{sec:rad-rec}
We estimate the emission from reconnection-accelerated particles in the coronae of X-ray binaries, following \citet{beloborodov_radiative_2017,sironi_kinetic_2020,sironi_relativistic_2025}. We remain agnostic about the global geometry of the current sheet and focus on a single representative layer. We first summarize the physical parameters of the corona, then identify the characteristic energy scales of reconnection-accelerated particles, and finally estimate the resulting photon spectrum. In Section~\ref{sec:analytical}, we discuss how this spectrum is reprocessed within the corona, which we identify as the reconnection region itself.

\subsection{Plasma conditions in the corona}
\label{subsec:corona-params}
We consider a black hole of mass $M \sim 10\,M_\odot$; its gravitational radius is $r_g = GM/c^2 \simeq 1.5\times10^{6}\,\rm cm$, where $G$ is the gravitational constant and $c$ is the speed of light. The strength of the coronal magnetic field is a major source of uncertainty. We adopt the estimate of \citet{beloborodov_radiative_2017}, who assume that the magnetic pressure in the corona is comparable to the mid-plane pressure of the accretion disk:
\begin{equation}
    B \sim 10^8 \, \text{G} \left( \frac{M}{10 M_\odot} \right)^{-1/2}.
\end{equation}
We use the convention $Q_X=Q/10^X$ in cgs units; for example, $B_8=B/10^8\,\rm G$. We take the reconnection layer to have length $R\sim2\,r_g\simeq3\times10^{6}\,\rm cm$ and adopt a fiducial reconnection rate $\beta_\text{rec}=0.1$. The corresponding layer thickness is set by the size of the largest plasmoids, $w\sim\beta_\text{rec}R\simeq3 \times10^{5}\,\rm cm$ \citep{sironi_plasmoids_2016}.

The magnetic compactness is 
\begin{equation}
    \ell_B = \frac{\sigma_T U_B R}{m_e c^2}  \simeq 10^3 B_8^2 R_{6.5},
    \label{eq:lB}
\end{equation}
where $U_B = B^2/(8\pi)$ is the energy density of the reconnecting field, $\sigma_T$ is the Thomson cross-section, and $m_e$ is the electron mass. Note that if $B\propto M^{-1/2}$ and $R$ is scaled to $r_g\propto M$, the magnetic compactness is independent of BH mass.

We assume that the inflowing (upstream) plasma consists of pairs with total number density $n$. The composition of the BH corona depends on its location and geometry and remains poorly constrained \citep{sridhar_bulk_2025}. The hard-state spectrum is consistent with Comptonization in a plasma with Thomson optical depth $\tau_T\sim1$--$3$ \citep{1997MNRAS.288..958G,mcconnell_soft_2002}. We assume that, in our scenario, the Comptonizing plasma is the reconnected (downstream) plasma, with number density comparable to that of the upstream plasma. Adopting $\tau_T=n\sigma_T  R\sim1$ then provides an estimate for $n$ and for the cold magnetization:\footnote{Alternatively, one could assume that the optical depth across the layer is $n\sigma_T w\sim1$. With this choice, the estimated magnetization would decrease by a factor of $1/\beta_{\rm rec}$.} 
\begin{equation}\label{eq:scales-sigma}
    \sigma = \frac{B^2}{4 \pi m_e n c^2} \sim \frac{2 \ell_B}{\tau_T} \simeq 2 \times 10^3 B_8^2 R_{6.5}.
\end{equation}
 These parameters place reconnection in the relativistic ($\sigma \gg 1$) and radiative ($\ell_B \gg 1$) regime.

\subsection{Energy scales of the accelerated particles}\label{subsec:scales}
Dissipation of magnetic energy via reconnection imparts a characteristic Lorentz factor $\gamma\sim\sigma$ to particles accelerated at X-points. A ``free'' particle continuously accelerated by the reconnection electric field, $E_\text{rec}=\beta_\text{rec}B$, over the characteristic advection time, $t_\text{adv}\sim R/c$, would instead reach a maximum Lorentz factor
\begin{equation}\label{eq:gamma-max}
    \gamma_{\text{max}} = \frac{\beta_\text{rec} e B R}{m_e c^2} \simeq 2 \times 10^{10} B_8 R_{6.5},
\end{equation}
where $e$ is the elementary charge.

Radiative losses, however, prevent particles from reaching \gammamax{}. We define the radiation-reaction limit as the Lorentz factor at which the radiative drag force balances the accelerating force from the electric field, $\gamma_\text{rad} = \sqrt{3e \beta_\text{rec} B/(4\sigma_T U)}$, where $U=U_B$ for synchrotron emission (with the drag averaged over an isotropic distribution of pitch angles) and $U=U_\text{rad}$ for IC scattering off the ambient radiation field. The Lorentz factor \gammarad{} therefore sets the maximum Lorentz factor attainable by radiating particles. For synchrotron emission, however, this limit can be exceeded if the magnetic field component perpendicular to the particle motion in the acceleration region is $\ll B$, either because the local magnetic field is weak or because the particle moves nearly along the field (i.e., at a small pitch angle).

A particle with $\gamma \gg 1$ cools on a timescale $t_\text{cool} = 3R/(4\gamma \ell c)$, where $\ell = \sigma_T U R/(m_e c^2)$ is the compactness of the magnetic or radiation field \citep{beloborodov_radiative_2017}. The ratio of the cooling time to the advection time is $t_\text{cool}/t_\text{adv}=\tau_\text{cool}/\gamma$, where, following \citet{mehlhaff_pair_2021}, we define
\begin{equation}\label{eq:scales-tau-cool}
    \tau_{\text{cool}} = \frac{\gamma_{\text{rad}}^2}{\gamma_{\text{max}}} = \frac{3}{4 \ell}.
\end{equation}
In a region of high compactness, $\ell \gg 1$, we have $\tau_{\text{cool}} \ll 1$, so particles cool before escaping; this defines the fast-cooling regime. Since $\ell_B \gg 1$ (Eq.~(\ref{eq:lB})), the coronal plasma lies in this regime and radiates most of its energy before leaving the reconnection layer. Consequently, the electromagnetic energy flux entering the layer, $\sim \beta_\text{rec}cU_B$, is ultimately converted into radiation. For a moderate Thomson optical depth, this radiation escapes at $\sim c$, implying a radiation energy density $U_\text{rad}\sim\beta_\text{rec}U_B$ in the reconnection region. Since $U_\text{rad}<U_B$, we have $t_\text{cool}^\text{sync}<t_\text{cool}^\text{IC}$, making synchrotron emission the dominant cooling channel.
    
The synchrotron radiation-reaction limit of a particle with $\gamma \gg 1$ and pitch angle $\theta = \pi/2$ is
\begin{equation}
    \gamma_\text{rad}^\text{sync} = \sqrt{\frac{e \beta_\text{rec} B}{2 \sigma_T U_B}} \simeq 3\times 10^3 B_8^{-1/2},
\end{equation}
which gives $\tau_\text{cool}^\text{sync} \simeq 5 \times 10^{-4} B_8^{-2} R_{6.5}^{-1}$.

Synchrotron radiation from particles below the self-absorption threshold \gammassa{} is self-absorbed. \citet{beloborodov_radiative_2017} derive
\begin{equation}\label{eq:scales-ssa}
    \gamma_{\text{SSA}} \simeq 15 \left(\frac{M}{10 M_\odot} \right)^{1/12};
\end{equation}
we provide an independent estimate of this threshold in Section~\ref{sec:analytical}, finding a different mass scaling.

Below \gammassa{}, synchrotron cooling is balanced by SSA heating, so particles cool primarily through IC scattering off the ambient radiation. The IC radiation-reaction limit is
\begin{equation}
    \gamma_\text{rad}^{\text{IC}} = \sqrt{\frac{3 e \beta_\text{rec} B}{4 \sigma_T U_{\rm rad}}} \simeq 10^4 B_8^{-1/2},
\end{equation}
which results in $\tau_\text{cool}^\text{IC} \simeq 7 \times 10^{-3}B_8^{-2} R_{6.5}^{-1}$.

For these parameters, the characteristic energy scales in X-ray binary coronae follow the hierarchy:
\begin{equation}\label{eq:scales-hierarchy}
    \tau_\text{cool} \ll 1 < \gamma_\text{SSA} \ll \sigma < \gamma_\text{rad}^\text{sync} < \gamma_\text{rad}^\text{IC}.
\end{equation}
Since $\sigma < \gamma_\text{rad}^\text{sync}$, 3D acceleration of ``free'' particles becomes important: these particles can be accelerated beyond $\sigma$ and radiate high-energy synchrotron emission.

\subsection{Emission by the accelerated particles}\label{sec:rad-rec-emission}
\begin{figure}
    \centering
    \includegraphics[width=\columnwidth]{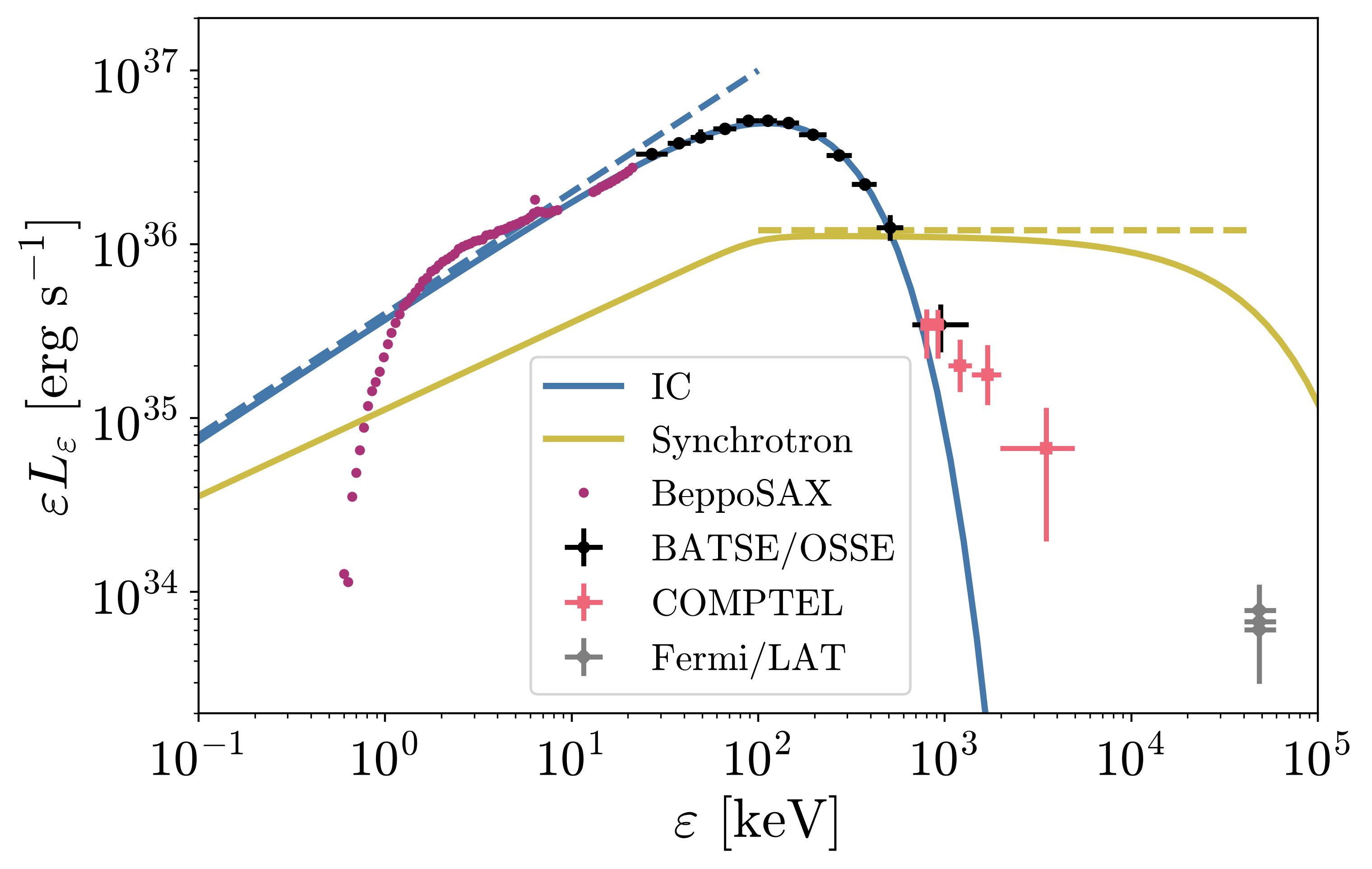}
    \caption{The IC and synchrotron emission spectra of reconnection-accelerated particles, together with observations of Cygnus~X-1 \citep{2001ApJ...547.1024D,mcconnell_soft_2002,2017MNRAS.471.3657Z}. Dashed lines show the power-law approximations used in Section~\ref{sec:analytical}. Effects of $\gamma \gamma$ pair production are not included here; see Sections~\ref{sec:analytical} and~\ref{sec:lehamoc}.}
    \label{fig:inj-spec}
\end{figure}

The hierarchy in Equation~(\ref{eq:scales-hierarchy}) divides the particles accelerated by reconnection into two populations: particles with $\gamma < \gamma_\text{SSA}$, which cool primarily through IC scattering, and particles with $\gamma > \gamma_\text{SSA}$, which cool primarily through synchrotron emission. We model the associated photon spectrum as a sum of the corresponding IC and synchrotron components:
\begin{equation}\label{eq:L-tot-inj}
    \varepsilon L_\varepsilon = \varepsilon L_\varepsilon^\text{IC} + \varepsilon L_\varepsilon^\text{sync} = L_1 g_1(\varepsilon) + L_2 g_2(\varepsilon),
\end{equation}
for $0.1 \, \text{keV} < \varepsilon < 100 \, \text{MeV}$, where $g_1$ and $g_2$ are dimensionless spectral shapes defined below, and $L_1$ and $L_2$ are their respective normalizations. We define the ratio of the total synchrotron and IC luminosities as
\begin{equation}
    \xi \equiv \frac{L_\text{sync}}{L_\text{IC}} = \frac{L_2 \int g_2(\varepsilon) d\ln\varepsilon}{L_1 \int g_1(\varepsilon)d \ln \varepsilon},
\end{equation}
which is constrained by particle-in-cell (PIC) simulations in Section~\ref{sec:pic}. In Section~\ref{sec:analytical}, we adopt $\xi=0.5$ as our fiducial value. The adopted photon spectrum (including both IC and synchrotron components) for $\xi=0.5$ is shown in Figure~\ref{fig:inj-spec}.

We assume that the IC component is powered by plasmoid-chain Comptonization \citep{beloborodov_radiative_2017}. We model the shape of the IC component as a power law with an exponential cutoff:
\begin{equation}\label{eq:inj-ic}
    g_1(\varepsilon) = \left( \frac{\varepsilon}{\varepsilon_1}\right)^a e^{-\varepsilon / {\varepsilon_1}}.
\end{equation}
We fit this component to the hard-state observations of Cygnus~X-1 above $1 \, \text{keV}$, where interstellar absorption is negligible \citep{2017MNRAS.471.3657Z}. We obtain $L_1 \simeq 1.3 \times 10^{37} \, \text{erg/s}$, $\varepsilon_1 \simeq 160 \, \text{keV}$, and $a \simeq 0.7$, adopting a distance of $d \simeq 2 \, \text{kpc}$ \citep{mcconnell_soft_2002}.

The corresponding IC luminosity is $L_\text{IC} \simeq 1.7 \times 10^{37} \, \text{erg/s}$. For $\xi=0.5$, the total radiated luminosity is therefore $L_\text{rad}=(1+\xi)L_\text{IC}\simeq2.5\times10^{37}\,\text{erg/s}$. This is consistent with the fast-cooling expectation that nearly all of the electromagnetic power entering the reconnection layer, $\sim  \beta_{\rm rec} cU_BR^2$, should be converted to radiation, as we demonstrate in Section~\ref{sec:pic}.

We describe the distribution of reconnection-accelerated particles above the SSA threshold as a broken power law, $dN/d\gamma \propto \gamma^{-s}$, with a break at $\gamma_\text{br} \sim 0.2\,\sigma$ \citep{Comisso2023} and slopes $s_< = 1$ below the break and $s_> = 2$ above it \citep{sironi_relativistic_2025, Bowyer2026}. The latter slope in 3D relativistic reconnection is found to be independent of $\sigma$ for $\sigma\gtrsim 3$; it is ultimately set by the dynamics of free particles \citep{zhang_origin_2023}. In the fast-cooling regime, the corresponding synchrotron spectrum is $\varepsilon L_\varepsilon \propto \varepsilon^\Gamma$ with slopes $\Gamma_< = 0.5$ and $\Gamma_> = 0$. We therefore model the synchrotron component as a broken power law (with smoothing parameter $k=0.2$) followed by an exponential cutoff:
\begin{equation}\label{eq:inj-sync}
    g_2(\varepsilon) = \left(\frac{\varepsilon}{\varepsilon_{\mathrm{br}}}\right)^{\Gamma_<}
    \left(1+\left(\frac{\varepsilon}{\varepsilon_{\mathrm{br}}}\right)^{1/k}\right)^{(\Gamma_>-\Gamma_<)k} 
    e^{-\varepsilon/\varepsilon_2}.
\end{equation}
The characteristic synchrotron energy of particles at the break Lorentz factor $\gamma_{\rm br} \sim 0.2\,\sigma$ is $\varepsilon \sim 300\,\text{keV}$, comparable to the peak of the IC component. We adopt $\varepsilon_\text{br}=100\,\text{keV}$, and have verified that order-unity variations around this value do not affect our conclusions (Appendix~\ref{app:lehamoc-sync}). Particles reaching the burnoff limit \gammasync{} emit photons near the synchrotron burnoff energy, $\sim 3 h e^2 \beta_\text{rec}/(\sigma_T m_e c) \sim 16$~MeV, which is independent of $B$ and depends only on $\beta_\text{rec}$ and fundamental constants ($h$ is the Planck constant). Because particles can moderately exceed \gammasync{} (e.g., due to small pitch angles), and because the reconnection rate at high magnetization may be larger than our fiducial $\beta_\text{rec}=0.1$ \citep{sironi_plasmoids_2016}, we adopt a fiducial synchrotron cutoff energy $\varepsilon_2= 45\,\text{MeV}$. In Appendix~\ref{app:lehamoc-sync}, we show that order-unity variations in the parameters of the synchrotron component do not qualitatively change our conclusions.

\section{Pair production in the corona}\label{sec:analytical}
In this section, we analytically estimate how pair production reprocesses the photon spectrum injected by reconnection-accelerated pairs. We adopt a homogeneous one-zone model in which the reconnection region is represented as a sphere of radius $R$, leaving anisotropy and spatial inhomogeneity to future work. We approximate the injected IC and synchrotron components, described in Section~\ref{sec:rad-rec-emission}, as power laws and assume that both photon populations are uniformly distributed throughout the reconnection region. We show that the $\gtrsim 1$~MeV synchrotron photons undergo $\gamma\gamma$ pair production with the IC photons, steepening the escaping $\gamma$-ray spectrum. The resulting secondary pairs can provide a Thomson optical depth of order unity, suggesting that they may constitute most (or all) of the pairs in the reconnection region. As they cool, the secondary pairs produce synchrotron emission that can serve as seed photons for Comptonization. We test these estimates numerically in Section~\ref{sec:lehamoc}.

\subsection{Power-law approximation}
To simplify the analytical calculations, we approximate  the spectral components described in Section~\ref{sec:rad-rec-emission} with power laws (Figure~\ref{fig:inj-spec}). The IC emission is then
\begin{equation}\label{eq:ic-power-law}
    \varepsilon L_\varepsilon^\text{IC} = L_1 (\varepsilon/\varepsilon_1)^a, \qquad \varepsilon_\text{min} \le \varepsilon \le \varepsilon_1,
\end{equation}
with fiducial parameters $a=0.7$, $L_1 = 10^{37}$~erg/s, $\varepsilon_\text{min} = 0.1$~keV, and $\varepsilon_1 = 100$~keV. The synchrotron $\gamma$-ray spectrum above $\varepsilon_{\rm br}$ is modeled as a power law
\begin{equation}
    \varepsilon L_\varepsilon^\text{sync} = L_2 (\varepsilon/\varepsilon_2)^b, \qquad \varepsilon_\text{br} \le \varepsilon \le \varepsilon_2,
\end{equation}
with fiducial values of $\varepsilon_\text{br} = 100$~keV, $\varepsilon_2 = 45$~MeV, and $b=0$. We adopt a synchrotron-to-IC power ratio $\xi = 0.5$ (supported by the PIC simulations in Section~\ref{sec:pic}), which results in
\begin{equation}\label{eq:L2-xi-approx}
L_2 = \frac{\xi}{a} L_1 \frac{1}{\Lambda} \simeq 1.2 \times 10^{36} \, \text{erg/s},
\end{equation}
where we have defined $\Lambda \equiv \ln(\varepsilon_2/\varepsilon_\text{br})$ for $b = 0$ and $\Lambda \equiv \left( 1 - (\varepsilon_\text{br}/\varepsilon_2)^b \right) / b$ otherwise.

We define the compactness of the IC component as 
\begin{equation}
    \ell_C = \frac{\sigma_T L_1}{4 \pi R m_e c^3} \simeq 7 \, B_8^2 R_{6.5},
\end{equation}
while the compactness of the synchrotron component is
\begin{equation} 
\ell_S = \frac{\sigma_T L_2}{4 \pi R m_e c^3} \simeq 0.8 \, B_8^2 R_{6.5}.
\label{eq:lS}
\end{equation}

\subsection{Optical depth of $\gamma \gamma \rightarrow e^{-}e^{+}$  interactions}\label{sec:analytical-tau}
Synchrotron $\gamma$-ray photons with energies above $m_e c^2$ can interact with ambient photons and produce secondary pairs. The most energetic $\gamma$-rays, with energy $\varepsilon_2=45$~MeV near the synchrotron burnoff limit, preferentially interact with target photons of energy $\varepsilon \gtrsim (m_e c^2)^2/\varepsilon_2 \simeq 5.8~\rm keV$, which are supplied by the IC component (Figure~\ref{fig:inj-spec}). We assume that IC photons provide the targets for all synchrotron $\gamma$-rays. We now estimate the optical depth to $\gamma\gamma$ pair production.

For a system with $\tau_T\lesssim 1$, the differential number density of IC photons is
\begin{equation}
    n^{\rm IC}_\varepsilon = \frac{L^{\rm IC }_\varepsilon t_{\rm adv}}{\varepsilon V} = n_\varepsilon^0 \left( \frac{\varepsilon}{\varepsilon_1} \right)^{a-2},
\end{equation}
where $n_\varepsilon^0 = 3 L_1 / (4 \pi R^2 c\, \varepsilon_1^2)$ and $V=4\pi R^3/3$. We now consider the interaction of a $\gamma$-ray photon of energy $\varepsilon_\gamma$ with a target photon of energy $\varepsilon$. We approximate the cross-section of $\gamma \gamma$ pair production as\footnote{Physically, $\sqrt{\zeta/2}$ is the Lorentz factor of the produced pairs in the center-of-momentum frame.}
\begin{equation}
    \sigma_{\gamma \gamma}(\zeta) \simeq \frac{\sigma_{\gamma \gamma,0}}{\zeta} \quad \text{for} \quad \zeta>2,
\end{equation} 
where $\zeta = \varepsilon \varepsilon_\gamma (1-\cos\chi)/(m_e c^2)^2$, $\chi$ is the collision angle of the two photons, and $\sigma_{\gamma \gamma,0} = 3\sigma_T/4$. This approximation is valid for $\zeta\gg2$. Closer to threshold, logarithmic corrections become important (see Eq.~10.4 in \citealt{2009herb.book.....D}).

For isotropic IC photons, the optical depth for $\gamma \gamma$ attenuation of a $\gamma$-ray photon is given by
\begin{equation}
    \tau_{\gamma \gamma}(\varepsilon_\gamma) = R\int d\varepsilon \frac{n^{\rm IC}_\varepsilon}{4 \pi}\int d\Omega (1-\cos\chi)\sigma_{\gamma \gamma}(\zeta),
\end{equation}
where the rightmost integral is over the directions of target photons. Our derivation is similar to that of \citet{1999MNRAS.305..181B}, except that they primarily considered $\sim\,$MeV photons interacting with each other, whereas here the target photons, produced by Comptonization and extending up to $\sim 100\,$keV, have a different origin from the $\gamma$-rays, which are produced by synchrotron emission and reach energies of tens of MeV.
\citet{1999MNRAS.305..181B} also accounted for anisotropies in the $\gamma\gamma$ interactions, which we neglect in our isotropic, homogeneous one-zone model. For anisotropic target photons, the opacity may differ by a factor of a few (Appendix~C of \citealt{2026ApJ..1000L..28P}).
 
In the asymptotic limit $\varepsilon_\gamma  \gg (m_e c^2)^2/\varepsilon_1 \simeq 2.6$~MeV, the optical depth becomes
\begin{align}
    \tau_{\gamma \gamma} &\simeq \frac{R\sigma_{\gamma \gamma,0} n_\varepsilon^0 \varepsilon_1}{2-a}\left(\frac{\varepsilon_\gamma \varepsilon_1}{m_e^2 c^4} \right)^{1-a} \\ &\simeq 40 \, \left(\frac{\varepsilon_1}{100\, {\rm keV}} \right)^{-0.7} \left( \frac{\varepsilon_\gamma}{m_e c^2} \right)^{0.3}.
\end{align}
The corona is therefore optically thick to $\gamma\gamma$ pair production for $\gamma$-rays above $\sim 2.6$~MeV, with the opacity increasing toward higher energies.

\subsection{Escaping $\gamma$-ray spectrum}\label{sec:analytical-tail}
To estimate the escaping $\gamma$-ray spectrum, we solve the kinetic equation for the number of $\gamma$-ray photons in the reconnection region, $N_\varepsilon^\gamma$:
\begin{equation}
    \frac{\partial N^\gamma_\varepsilon}{\partial t}  = Q_{\rm inj} + \mathcal{L}_{\gamma \gamma} + \mathcal{L}_{\rm esc},
\end{equation}
where $Q_{\rm inj} = \varepsilon L^{\rm sync}_{\varepsilon}/\varepsilon^2$ is the injection rate of synchrotron photons of energy $\varepsilon$ (in this subsection, we drop the subscript in $\varepsilon_\gamma $), $\mathcal{L}_{\gamma \gamma} = - N^\gamma_{\varepsilon} R_{\gamma \gamma}$ is the loss term due to $\gamma \gamma$ absorption with attenuation rate $R_{\gamma \gamma} = \tau_{\gamma \gamma} /t_{\rm adv} $, and $\mathcal{L}_{\rm esc} = - N^\gamma_{\varepsilon} / t_{\rm adv}$ is the escape rate. Since $\tau_{\gamma \gamma} \gg 1$, $\gamma \gamma$ attenuation dominates over escape, and the steady-state photon distribution is $N^\gamma_\varepsilon= Q_{\rm inj}/R_{\gamma \gamma}$. The escaping $\gamma$-ray spectrum is obtained as
\begin{equation}
    \varepsilon L^{\gamma}_{\varepsilon} = \frac{\varepsilon^2 N^\gamma_{\varepsilon}}{t_{\rm adv}} = \frac{\varepsilon L^{\rm sync}_{\varepsilon}}{\tau_{\gamma \gamma}(\varepsilon)}.
\end{equation}
Substituting the asymptotic expression for $\tau_{\gamma \gamma}$, valid for $\varepsilon \gg 2.6$~MeV, gives
\begin{equation}\label{eq:Lg-ss}
\varepsilon L^{\gamma}_{\varepsilon} \simeq  L_0^\gamma \left(\frac{\varepsilon}{m_e c^2}\right)^{-1+a+b},
\end{equation}
where
\begin{equation}
L_0^\gamma = \xi\frac{4\pi R m_e c^3}{3 \sigma_{\gamma \gamma,0}}\frac{2-a}{a} \frac{1}{\Lambda}\left(\frac{\varepsilon_1}{m_e c^2}\right)^a\left( \frac{\varepsilon_2}{m_e c^2} \right)^{-b}.
\end{equation}
While the injected synchrotron spectrum has slope $b$, the escaping spectrum has slope $-1+a+b$, as shown in Equation~(\ref{eq:Lg-ss}). Thus, $\gamma\gamma$ attenuation steepens the spectrum by $\Delta\Gamma=1-a$, in agreement with the general result derived by \citet{2013A&A...557A..48P} (see also their Appendix~A). We argue that the escaping $\gamma$-ray photons are a promising explanation for the MeV tail of X-ray binaries, potentially extending to several MeV. In Section~\ref{sec:lehamoc}, we show that the shape and normalization of the escaping spectrum are in good agreement with observations.

\subsection{Secondary pair distribution}\label{subsec:sec-dist}
We now estimate the energy distribution of the secondary pairs and the resulting Thomson optical depth. The number of secondary pairs per unit Lorentz factor $\gamma$, which we denote as $N^\pm_\gamma$, evolves as
\begin{equation}\label{eq:N-pm-gamma}
    \frac{\partial N^\pm_\gamma} {\partial t}  = \mathcal{L}^\pm_{\rm esc} +  \mathcal{L}^\pm_{\rm sync} + \mathcal{Q}^\pm_{\gamma \gamma},
\end{equation}
where $\mathcal{L}^\pm_{\rm esc} = - N^\pm_\gamma / t_{\rm adv}$ is the pair escape rate, $\mathcal{L}^\pm_{\rm sync} = \partial(\dot{\gamma}_{\rm sync} N^\pm_\gamma) / \partial \gamma$ describes synchrotron losses (which dominate above the SSA threshold) with cooling rate $\dot{\gamma}_{\rm sync} = \gamma / t_\text{cool}^\text{sync}$, and $\mathcal{Q}^\pm_{\gamma \gamma} = - 4 m_e c^2\mathcal{L}_{\gamma \gamma}(\varepsilon_\gamma)$ is the pair production rate evaluated at $\varepsilon_\gamma = 2 \gamma m_e c^2$. The factor of four in $\mathcal{Q}^\pm_{\gamma \gamma}$ arises because a $\gamma$-ray photon produces two particles, each carrying half of the photon energy.

Using $\mathcal{Q}^\pm_{\gamma \gamma} = 4 m_e c^2 Q_{\rm inj}$ (Section~\ref{sec:analytical-tail}), and assuming that synchrotron losses dominate over escape, we obtain the volume-integrated steady-state distribution of secondary pairs:
\begin{equation}
N^\pm_\gamma \simeq \frac{1}{1 - b} \frac{3\pi R^2}{\sigma_T} \frac{\ell_S}{\ell_B} \left(\gamma^\pm_{\max}\right)^{-b} \frac{\gamma^{-1+b}}{\gamma^2-1}, 
\label{eq:dndg}
\end{equation}
where $\gamma^\pm_{\max} = \varepsilon_2 / (2 m_e c^2) \simeq 44 \, (\varepsilon_2 / 45~\text{MeV})$ is the maximum Lorentz factor of secondary pairs set by the highest-energy synchrotron photons. For $\gamma \gg 1$, the distribution is soft, $N^\pm_\gamma \propto \gamma^{-3+b}$, i.e., most of the pairs reside at low Lorentz factors.

In the fast-cooling regime, the secondary pairs cool down to the Lorentz factor at which the cooling and escape timescales become comparable. We can then estimate the number density of secondary pairs in the reconnection region $n^\pm = N^\pm / V$ by balancing injection $\sim L_2 / (m_e c^2)$ and escape $\sim N^\pm / t_\text{adv}$. Their Thomson optical depth is
\begin{equation}\label{eq:tau-t-pm}
    \tau_T^\pm = n^\pm \sigma_T R \sim \frac{L_2 t_\text{adv}}{m_e c^2 V} \sigma_T R \sim 3 \ell_S \simeq 2\, B_8^2 R_{6.5}.
\end{equation}
The same estimate can be obtained by integrating the distribution in Equation~(\ref{eq:dndg}) down to the Lorentz factor at which cooling and escape become comparable. It follows that secondary pairs, produced by $\gamma\gamma$ interactions involving high-energy synchrotron photons, can account for most, if not all, of the pairs in the reconnection region. The system can therefore self-regulate: pairs injected into the upstream plasma through $\gamma\gamma$ pair production are subsequently processed and accelerated by reconnection, generating the high-energy synchrotron photons that sustain pair production in the first place.

Our estimate of the secondary-pair density neglects pair annihilation. For non-relativistic pairs, the ratio of the annihilation rate, evaluated using the pair density in Equation~(\ref{eq:tau-t-pm}), which neglects annihilation, to the escape rate is \citep[e.g.,][]{Coppi1990,beloborodov_radiative_2017}
\begin{equation}
    \frac{\mathcal{L}^\pm_{ee}}{N^\pm/t_{\rm adv}} \sim \frac{9 \ell_S}{8} \simeq 0.9 B_8^2 R_{6.5},
\end{equation}
showing that annihilation is unlikely to modify our estimate of $\tau_T^\pm$ (Eq.~(\ref{eq:tau-t-pm})) by more than a factor of order unity. Similarly, the luminosity of the annihilation line at $\varepsilon \sim m_e c^2$ is not expected to significantly exceed the synchrotron luminosity, 
\begin{equation}
\frac{\mathcal{L}_{ee}^\pm m_e c^2}{L_2} \sim \frac{9}{8} \ell_S \simeq 0.9 B_8^2 R_{6.5}.
\end{equation}

\subsection{Synchrotron emission by the secondary pairs}\label{sec:analytical-pm-spec}
We assume that secondary pairs cool before being re-accelerated by reconnection. We adopt a monoenergetic approximation for their synchrotron emissivity, $j_{\rm sync}(\nu, \gamma)= (4/3) \sigma_T U_B c \gamma^2 \delta(\nu - \nu_{\rm sync})$, where $\nu_{\rm sync} = eB\gamma^2/(2\pi m_e c)$ is the characteristic frequency emitted by particles of Lorentz factor $\gamma$. In the optically thin limit, the synchrotron spectrum of the cooling pairs follows from Eq.~(\ref{eq:dndg}):
\begin{equation}\label{eq:sec-sync-spec}
    \varepsilon L_\varepsilon^\pm \simeq \frac{4}{3}\sigma_T U_B c \gamma^3 N_\gamma^\pm \simeq L_2.
\end{equation}
The last step assumes $b=0$ and shows that the synchrotron emission from secondary pairs has approximately the same luminosity per logarithmic photon-energy interval as the injected synchrotron component. Their spectrum extends up to the synchrotron photon energy emitted by secondary pairs with the maximum Lorentz factor $\gamma_{\rm max}^\pm$, which is set by the burnoff limit of the injected synchrotron photons, as discussed above. We therefore expect a cutoff at $\varepsilon_{\rm max}^\pm \simeq 2.2\, (\varepsilon_2/45\,\mathrm{MeV})^2 B_8\,\mathrm{keV}$. 

At sufficiently low frequencies, however, this emission becomes self-absorbed. The optical depth to synchrotron self-absorption is $\tau_{\nu, \rm SSA}=a_{\nu, \rm SSA} R$, where the absorption coefficient is \citep{1986rpa..book.....R}
\begin{equation}\label{eq:a-nu-ssa}
    a_{\nu, \rm SSA} = -\frac{c^2}{8\pi V \nu^2 m_e c^2}\int {\rm d}\gamma \, \gamma^2 \frac{\partial}{\partial \gamma}\left( \frac{N^\pm_\gamma}{\gamma^2}\right)j_{\rm sync}(\nu, \gamma).
\end{equation}
Substituting the secondary-pair distribution (Eq.~(\ref{eq:dndg})) and setting $\tau_{\nu, \rm SSA}=1$ gives the SSA photon energy:
\begin{equation}\label{eq:eps-ssa}
    \varepsilon_{\rm SSA} = \left(\frac{15 h^3 c^3 \ell_S}{16\pi\sigma_T R}\right)^{2/7}\left(\frac{heB}{2\pi m_e c}\right)^{1/7} \simeq 35 \, B_8^{5/7}~\rm eV.
\end{equation}
At photon energies $\varepsilon<\varepsilon_{\rm SSA}$, the synchrotron spectrum of secondary pairs rises as $\varepsilon L_\varepsilon^\pm\propto\varepsilon^{7/2}$, while above $\varepsilon_{\rm SSA}$ it flattens (Eq.~(\ref{eq:sec-sync-spec})). The secondary synchrotron emission therefore peaks near $\varepsilon_{\rm SSA}$. The Lorentz factor of particles emitting at $\varepsilon_{\rm SSA}$ is
\begin{equation}
    \gamma_\text{SSA} \simeq 5\, B_8^{-1/7} \simeq 5 \left( \frac{M}{10 \, M_\odot} \right)^{1/14}.
\end{equation}
Thus, $\gamma_{\rm SSA}$ depends only weakly on the magnetic field strength and BH mass. Note that the dependence on BH mass differs slightly from the $M^{1/12}$ scaling derived by \citet{beloborodov_radiative_2017}, who assumed an energy-independent pair yield.

We conclude by assessing whether synchrotron photons emitted by the cooling secondary pairs can serve as seed photons for Comptonization. We compare the luminosity at the peak  ($\sim \varepsilon_\text{SSA}$) of the secondary synchrotron spectrum, $\varepsilon L_\varepsilon^\pm$, with that of the IC component, $\varepsilon L_\varepsilon^\text{IC}$, at the same photon energy. For secondary synchrotron photons to provide a significant source of seed photons, the former must exceed the latter. Combining Equations~(\ref{eq:ic-power-law}), (\ref{eq:L2-xi-approx}), and (\ref{eq:sec-sync-spec}), we find that this requires
\begin{equation}\label{eq:xi-seed-photons}
    \xi > \xi_s \simeq a \Lambda \left( \frac{\varepsilon_\text{SSA}}{\varepsilon_1} \right)^a \simeq 0.02 \, B_8^{1/2}.
\end{equation}
We test this prediction numerically in Section~\ref{sec:lehamoc}.

\section{Energy partition in reconnection}\label{sec:pic}
In Sections~\ref{sec:rad-rec} and \ref{sec:analytical}, we have assumed fiducial values for the radiative efficiency of reconnection and for the ratio of synchrotron to IC power, $\xi$. To determine these quantities, we carry out 3D radiative particle-in-cell simulations of relativistic magnetic reconnection. We study the energization of reconnection-accelerated pairs and quantify how the dissipated magnetic energy is partitioned between IC and synchrotron emission. 

\subsection{PIC setup}
We perform a suite of 3D radiative kinetic simulations of relativistic reconnection in electron--positron plasma using the particle-in-cell (PIC) code \texttt{Tristan v2} on a uniform Cartesian grid \citep{tristan}. We do not include pair production; we treat it separately in Section~\ref{sec:lehamoc} by means of dedicated radiative calculations. Our initial conditions resemble those of the non-radiative 3D reconnection simulations by \citet{zhang_fast_2021}; we describe our cooling prescription below. We initialize a Harris sheet with the reversing magnetic field parameterized by $\bm{B} = B \, \tanh{(2 \pi y / \Delta)} \, \bm{\hat{x}}$, where $B$ is the initial magnetic field strength and $\Delta$ is the thickness of the current sheet. We also prescribe a field component in the $z$-direction, which we refer to as a (weak) guide field, $\bm{B}_g = 0.1 \, B \, \bm{\hat{z}}$. We choose the cold magnetization to be $\sigma = 10$, which allows us to approach the $\sigma \gg 1$ regime relevant to BH coronae (see Section~\ref{subsec:corona-params}).

The domain sizes are $R_x = 1638 \, c/\omega_p$ along the outflow, $R_y = 600 \, c/\omega_p$ along the inflow, and $R_z = 816 \, c/\omega_p$ along the guide field, where $c/\omega_p\equiv\sqrt{m_ec^2/(4\pi e^2 n_0)}$ is the background plasma skin depth. For this box, the maximum attainable Lorentz factor in Eq.~(\ref{eq:gamma-max}) is $\gamma_\text{max} \sim 260$. We additionally perform a set of simulations with smaller domains, described in Appendix~\ref{app:2k}. We define the characteristic system size as the half-length of the domain along the outflow direction, $R \equiv R_x / 2$.

We advance the positions and momenta of the particles using the Vay particle pusher \citep{vay_simulation_2008}. We have found it to be superior to the standard Boris algorithm in maintaining the outflow direction consistently along the $x$-axis; in the most strongly cooled cases, the Boris pusher instead produces an artificial tilt of the layer if the spatial resolution is insufficient. The numerical speed of light is $c = 0.45 \, \Delta x / \Delta t$ and the plasma skin depth is $c/\omega_p = 2.5 \, \Delta x$, where $\Delta x$ and $\Delta t$ are the cell size and timestep, respectively. We initialize the background plasma with a total number density (including both species) of $n_0 = 2 / (\Delta x)^3$.

We trigger reconnection by removing the plasma pressure near the center of the layer at $t=0$, which generates two reconnection fronts moving toward the $x$-boundaries at nearly the speed of light. We use periodic boundary conditions in $z$. The boundary conditions in $x$ are initially periodic and switch to outflow at $t \simeq 0.7 \, R/c$, when the two fronts approach the boundaries. At the $y$-boundaries, particles can exit freely, fields are damped to their initial values, and fresh magnetized plasma is continuously injected. The combination of outflow boundaries in $x$ and inflow boundaries in $y$ allows the system to reach a steady state after $t_i = 3 \, R/c$. We evolve the simulations up to $t_f = 10 \, R/c$ and time-average quantities between $t_i$ and $t_f$.

\subsection{Radiative cooling prescription}\label{sec:pic-cool}
\begin{figure}
  \centering
  \begingroup
\def\gscale{1.0}
\def\yaspect{0.5}  

\def\xtauIC{0.45}   
\def\xone{1.1}      
\def\xtausync{1.95} 
\def\xSSA{2.8}      
\def\xhw{0.15}      
\def\xIC{4.0}       
\def\xsigma{5.3}    
\def\xsync{7.55}     
\def\xcut{\xsync}   
\def\cutsharp{1.0}  
\def\xfree{\xsigma} 
\def\xstart{\xone}  
\def\xend{8.6}      

\def\dytauIC{0pt}
\def\dyone{0pt}
\def\dytausync{0pt}
\def\dySSA{0pt}
\def\dyIC{0pt}
\def\dysig{0pt}
\def\dysync{0pt}

\def\pone{1.0}      
\def\ptwo{3.0}      

\def\ystart{4.6}    
\def\ytop{6.2}      

\def\braceamp{5pt}  
\def\bracegap{3pt}  

\colorlet{ICcol}{blue!60!black}
\colorlet{SYNcol}{orange!85!black}

\pgfmathsetlengthmacro{\unit}{\gscale*\columnwidth/(\xend+1.2)}
\pgfmathsetmacro{\slopeA}{\pone-1}   
\pgfmathsetmacro{\slopeB}{\ptwo-1}   
\pgfmathsetmacro{\cutk}{2.302585*\cutsharp}

\begin{tikzpicture}[x=\unit, y={\yaspect*\unit}, >={Stealth[length=2.5mm]}]
  \fill[ICcol,  opacity=0.12] (0,0)               rectangle ({\xSSA+\xhw},\ytop);
  \fill[SYNcol, opacity=0.14] ({\xSSA-\xhw},0)    rectangle (\xend,\ytop);

  \node[ICcol,  anchor=north, yshift=-3pt] at ({(\xSSA-\xhw)/2},\ytop)        {IC-cooled};
  \node[SYNcol, anchor=north, yshift=-3pt] at ({(\xSSA+\xhw+\xend)/2},\ytop)  {synchrotron-cooled};

  \draw[gray, densely dotted] (\xIC,0)   -- (\xIC,\ytop);
  \draw[gray, densely dotted] (\xsync,0) -- (\xsync,\ytop);
  \draw[black!70, dashed, thick] (\xSSA,0) -- (\xSSA,\ytop);

  \begin{scope}
    \clip (0,0) rectangle (\xend,\ytop);
    \draw[very thick, black, samples=400, domain=\xstart:\xend]
      plot (\x, {\ystart
                 - \slopeA*(min(\x,\xsigma)-\xstart)
                 - \slopeB*max(\x-\xsigma,0)
                 - 0.4343*exp(\cutk*(\x-\xcut))});
  \end{scope}

  \draw[decorate, decoration={brace, amplitude=\braceamp}, thick]
    ([yshift=\bracegap]\xfree,\ytop) -- ([yshift=\bracegap]\xend,\ytop)
    node[midway, above=\braceamp+2pt] {``Free'' particles};

  \draw[->, thick] (0,0) -- (\xend+0.3,0);
  \draw[->, thick] (0,0) -- (0,\ytop+0.3);

  \node[anchor=north east] at (\xend+0.3,0) {$\gamma$};
  \node[rotate=90, anchor=south east, xshift=-1pt]
    at (0,\ytop-0.5) {$\gamma\,dN/d\gamma$};

  \foreach \xx/\lab/\dy in {\xtauIC/{\tau_{\mathrm{cool}}^{\mathrm{IC}}}/\dytauIC,
                            \xone/{1}/\dyone,
                            \xtausync/{\tau_{\mathrm{cool}}^{\mathrm{sync}}}/\dytausync,
                            \xSSA/{\gamma_{\mathrm{SSA}}}/\dySSA,
                            \xIC/{\gamma_{\mathrm{rad}}^{\mathrm{IC}}}/\dyIC,
                            \xsigma/{\sigma}/\dysig,
                            \xsync/{\gamma_{\mathrm{rad}}^{\mathrm{sync}}}/\dysync} {
    \draw[thick] (\xx,0) ++(0,-2pt) -- ++(0,4pt);
    \node[anchor=base, yshift=-13pt-\dy] at (\xx,0) {$\lab$};
  }

\end{tikzpicture}
\endgroup
  \caption{Schematic of the particle energy distribution, $\gamma \, dN / d \gamma$. Shaded regions represent dominant cooling mechanisms: IC (blue) or synchrotron (orange). The characteristic energy scales are ordered as adopted in the PIC simulations; the hierarchy expected in BH coronae is given by Eq.~(\ref{eq:scales-hierarchy}). }
  \label{fig:cooling-schematics}
\end{figure}
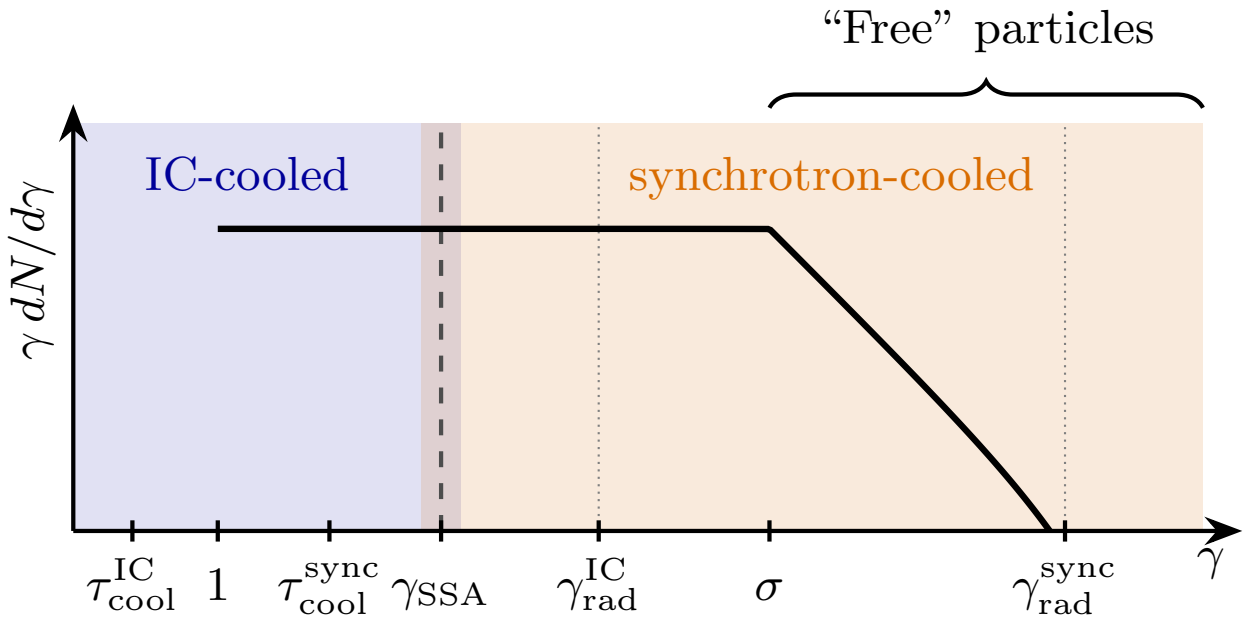
Figure~\ref{fig:cooling-schematics} illustrates our cooling prescription and the key scales of radiative reconnection adopted in our PIC simulations (Section~\ref{sec:rad-rec}). We model radiative losses due to IC and synchrotron cooling as continuous drag forces. Each force is characterized by its classical radiation-reaction limit, \gammaic{} and \gammasync{}, respectively:
\begin{align}
    \bm{F}_\text{IC} &= - e B \beta_\text{rec} \frac{\gamma^2}{\left( \gamma_\text{rad}^\text{IC} \right)^2} \bm{\beta}, \\
    \bm{F}_\text{sync} &= e B \frac{\beta_\text{rec}}{\left( \gamma_\text{rad}^\text{sync}\right)^2} \left[ \bm{\kappa}_R - \gamma^2 \chi_R^2 \bm{\beta}  \right], \\
    \bm{\kappa}_R &= (\bm{e} + \bm{\beta} \times \bm{b}) \times \bm{b} + (\bm{\beta} \cdot \bm{e}) \bm{e}, \\
    \chi^2_R &= (\bm{e} + \bm{\beta} \times \bm{b})^2 - (\bm{\beta} \cdot \bm{e})^2,
\end{align}
where $\beta_\text{rec} = 0.1$ as adopted in Sections~\ref{sec:rad-rec} and \ref{sec:analytical}, $\bm{\beta}$ is the particle dimensionless three-velocity (corresponding to Lorentz factor $\gamma$), and $\bm{e} = \bm{E} / B$ and $\bm{b} = \bm{B} / B$ are dimensionless fields \citep{landau_classical_1975}.

As discussed in Section~\ref{sec:rad-rec}, the two cooling channels dominate at different particle energies: particles below \gammassa{} cool primarily through IC scattering, whereas those above \gammassa{} cool primarily through synchrotron emission. The effective drag forces in our simulations are parameterized as follows:
\begin{align}
     \bm{F}_\text{IC}^\text{eff} &= \bm{F}_\text{IC} \cdot \frac{1}{2} \left( 1 - \tanh{\frac{\gamma - \gamma_\text{SSA}}{\Delta_{\gamma} \cdot \gamma_\text{SSA}}} \right), \label{eq:eff-drag-force-ic}  \\
      \bm{F}_\text{sync}^\text{eff} &= \bm{F}_\text{sync} \cdot \frac{1}{2} \left( 1 + \tanh{\frac{\gamma - \gamma_\text{SSA}}{\Delta_{\gamma} \cdot \gamma_\text{SSA}}} \right), \label{eq:eff-drag-force-sync}
\end{align}
where $\Delta_{\gamma}$ controls the fractional width of the transition between the two cooling channels around \gammassa{}. We have verified that 2D simulations with $\Delta_{\gamma} = 0.1$ and $\Delta_{\gamma} = 0.05$ yield consistent reconnection rates and time-averaged particle spectra. We therefore adopt $\Delta_{\gamma} = 0.05$ for our 3D simulations. We explore $\gamma_\text{SSA} = 3$ and $5$, both satisfying the hierarchy $\gamma_\text{SSA} < \sigma$ in Eq.~(\ref{eq:scales-hierarchy}).

For IC cooling, we adopt $\gamma_\text{rad}^\text{IC} = 7, \, 10$, corresponding to $\tau_\text{cool}^\text{IC} = 0.2, \, 0.4$. In BH coronae, $\gamma_\text{rad}^\text{IC} \gg \sigma$; however, capturing the fast-cooling regime, $\tau_\text{cool}^\text{IC} < 1$, in our simulations requires choosing $\gamma_\text{rad}^\text{IC} \leq \sigma$. This choice does not significantly affect our results for two reasons. First, since \gammaic{}$\,>\,$\gammassa{}, synchrotron emission becomes the dominant cooling channel before particles reach the IC radiation-reaction limit. Thus, particles are largely insensitive to whether \gammaic{} lies above or below $\sigma$. Second, the simulations in Appendix~\ref{app:2k} suggest that the results depend on the strength of IC cooling only through \taucoolic{} (which is $\ll 1$ in our simulations, consistent with coronal conditions), rather than directly on $\gamma_\text{rad}^\text{IC}$.

For synchrotron cooling, we adopt $\gamma_\text{rad}^\text{sync} = 21, \, 30, \, 42$, ensuring that $\gamma_\text{rad}^\text{sync} > \sigma$, as expected in BH coronae, and that synchrotron cooling does not suppress the ``free'' phases of particle acceleration (see Section~\ref{sec:intro}). After being accelerated to $\gamma \gg \sigma$ during the ``free'' phase, particles are recaptured by flux ropes, where they cool. Whether they cool via synchrotron emission down to \gammassa{}, below which IC cooling takes over and removes their remaining energy, depends on the value of \taucoolsync{} relative to \gammassa{}. In BH coronae, we expect \taucoolsync{} $\ll 1<\gamma_{\rm SSA}$. In our simulations, however, $\tau_\text{cool}^\text{sync} = 1.7, \, 3.5, \, 6.8$, comparable to \gammassa{}. Consequently, some particles may escape before radiating as much synchrotron energy as they would for $\tau_\text{cool}^\text{sync} \ll \gamma_\text{SSA}$, making our estimates of the synchrotron power in Section~\ref{sec:pic-results} lower limits.

\begin{deluxetable}{ccccc}[t!]
\tablecaption{\label{tab:sim}Simulation parameters and measured reconnection rates.}
    \tablehead{
    \colhead{ID} & \colhead{\gammaic{}} & \colhead{\gammassa{}} & \colhead{\gammasync{}} & \colhead{\betarec{}}
    }
    \startdata
    \ic{} & $10$ & $-$ & $-$ & $0.08$ \\
    \sync{} & $-$ & $-$ & $30$ & $0.07$ \\
    \hybrid{} & $10$ & $5$ & $30$ & $0.07$ \\
    \hybridicstrong{} & $7$ & $3$ & $21$ & $0.08$ \\
    \hybridic{} & $7$ & $3$ & $30$ & $0.08$ \\
    \hybridsync{} & $10$ & $5$ & $42$ & $0.07$
    \enddata
    \tablecomments{ID of the simulations and the corresponding cooling parameters \gammaic{}, \gammassa{}, and \gammasync{}, as defined in Section~\ref{sec:rad-rec}, together with the measured reconnection rate \betarec{}. All runs use $\sigma = 10$ and a guide field $B_g = 0.1 \, B$.}
\end{deluxetable}

We label our simulations by the three energy scales: \gammaic{}, \gammassa{}, and \gammasync{}; for instance, \hybrid{} has $\gamma_\text{rad}^\text{IC} = 10$, $\gamma_\text{SSA} = 5$, and $\gamma_\text{rad}^\text{sync} = 30$. Runs that include only IC or only synchrotron losses are denoted \ic{} and \sync{}, respectively. Table~\ref{tab:sim} summarizes the PIC simulations in this paper.

\subsection{Results}\label{sec:pic-results}
\begin{figure}[!t]
    \centering
    \includegraphics[width=\columnwidth]{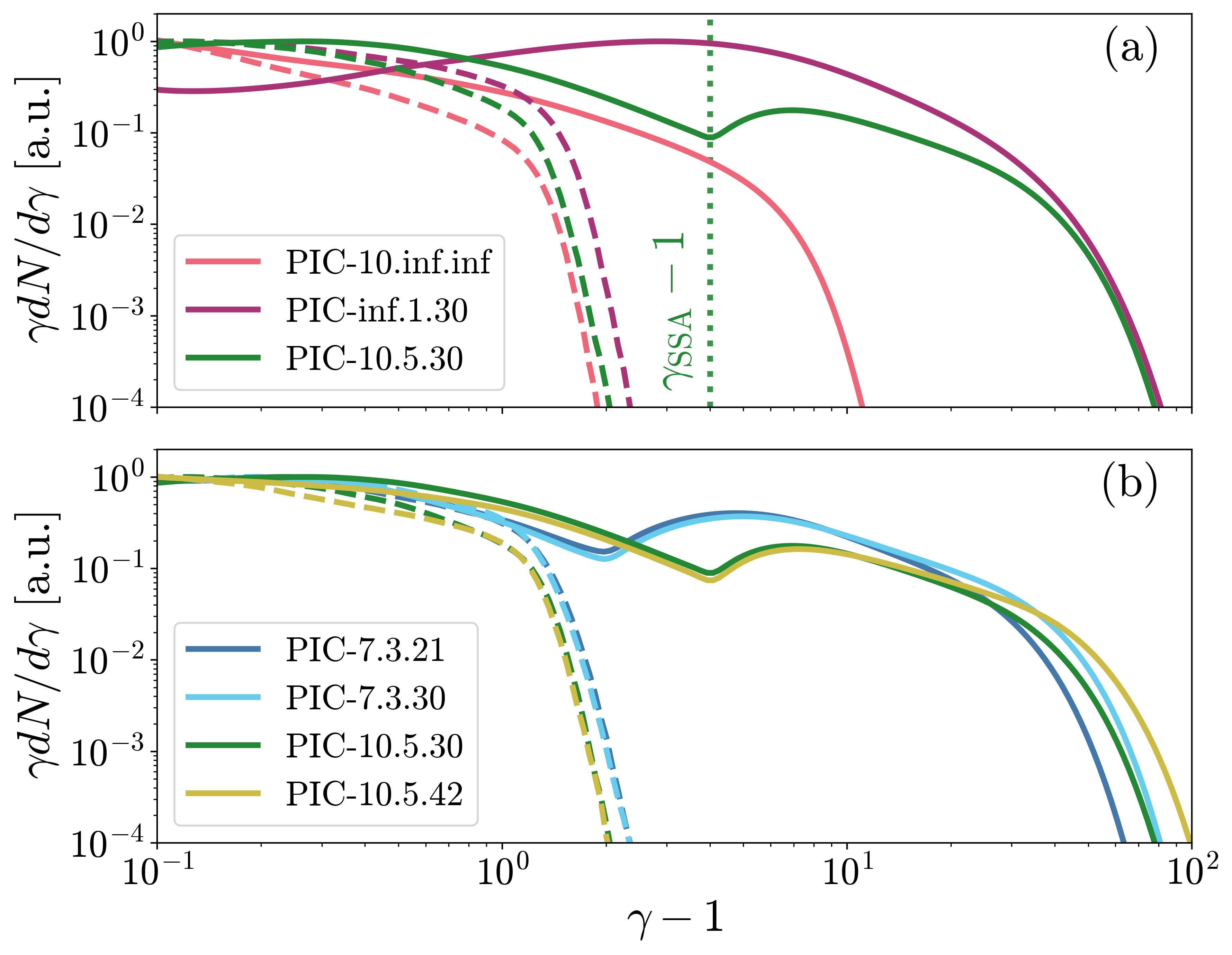}
    \caption{Time-averaged total-energy (solid) and bulk-energy (dashed) spectra. Panel (a) shows that in \hybrid{}, the particle spectrum follows that in \ic{} below the SSA threshold, $\gamma < \gamma_{\text{SSA}}$, and then tracks \sync{} at the highest energies. Panel (b) shows a comparison between simulations with different \gammaic{}, \gammassa{}, and \gammasync{}.}
    \label{fig:spec}
\end{figure}

Figure~\ref{fig:spec}(a) shows how IC and synchrotron cooling losses shape the particle energy distribution. Below the SSA threshold, $\gamma < \gamma_\text{SSA}$, the total-energy spectrum\footnote{We define ``total energy'' as the combination of bulk and internal components. Later we isolate the bulk contribution.} in \hybrid{} resembles that in the IC-only simulation, \ic{}, as expected given their identical \gammaic{} and \taucoolic{}. As particles approach and exceed the SSA threshold, synchrotron emission becomes the dominant cooling channel. The particle spectrum above $\gamma_\text{SSA}$ develops a peak around $\gamma - 1 \simeq 7$, followed by a power law, $ dN/d\gamma \propto \gamma^{-s}$, with $s \simeq 2.8$ for $\gamma - 1 \in [8, 50]$, similar to that in the synchrotron-only simulation, \sync{}. Previous non-radiative simulations of 3D reconnection demonstrated that the slope of $dN/d\gamma\propto\gamma^{-s_>}$ at $\gamma\gtrsim \sigma$ is nearly independent of magnetization, with $s_>=2$ \citep[][as assumed in Section~\ref{sec:rad-rec-emission}]{zhang_origin_2023}. Radiative cooling is expected to steepen this slope to $s_>+1=3$ \citep{Bowyer2026}, consistent with our measured value of $s\simeq2.8$. This corresponds to a nearly flat photon spectrum, with $\Gamma_>\simeq0$.

Figure~\ref{fig:spec}(b) shows how the particle energy spectrum depends on \gammaic{}, \gammassa{}, and \gammasync{}. Despite differences in cooling strength, all simulations exhibit qualitatively similar spectra: a low-energy peak largely shaped by bulk motions (compare with the dashed lines), a trough marking the transition to synchrotron-dominated cooling, a second peak at higher energies, and a power-law tail with a cutoff at the burnoff limit.

The low-energy end of the total-energy spectra is dominated by bulk motions (compare dashed and solid lines). The closest agreement between dashed and solid lines occurs in \hybridicstrong{} and \hybridic{}, the simulations with the lowest \gammaic{} (and \taucoolic{}), and hence the strongest IC cooling. We find that the spectrum of bulk motions is largely independent of \gammaic{}. In other words, Compton drag does not significantly suppress reconnection-driven bulk motions in this parameter regime.

\begin{figure}[!t]
    \centering
    \includegraphics[width=\columnwidth]{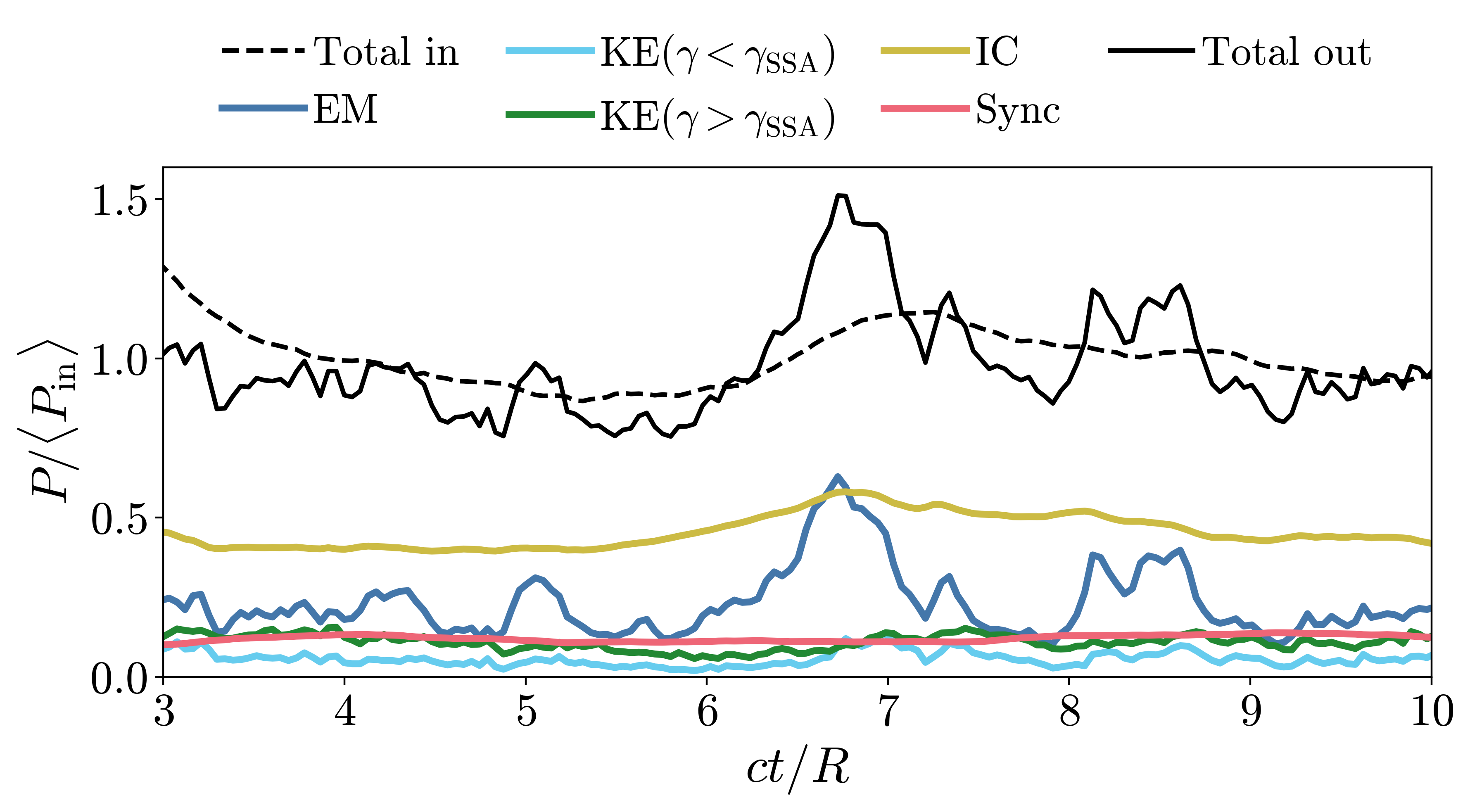}
    \caption{Temporal evolution of the total incoming and outgoing power in \hybrid{}, normalized by $\langle P_\text{in} \rangle$. The outgoing power is decomposed into low-frequency electromagnetic (EM), kinetic (KE), IC, and synchrotron (sync) components, with the kinetic component further split based on the energy of the escaping particles: below the SSA threshold, $\gamma < \gamma_\text{SSA}$, and above, $\gamma > \gamma_\text{SSA}$. The approximately constant $P_\text{in}/\langle P_\text{in}\rangle$ (dashed black) and $P_\text{out}/\langle P_\text{in}\rangle$  (solid black) indicate that the layer has reached a steady state.}
    \label{fig:power-time}
\end{figure}

We now examine how the incoming electromagnetic energy is partitioned among outgoing low-frequency electromagnetic energy, particle kinetic energy (including both bulk and internal motions), and radiation. We consider the volume enclosed by $x = \pm 0.9 \, R$ and $y = \pm 0.25 \, R$, excluding the regions where boundary conditions are imposed. The incoming (outgoing) low-frequency electromagnetic power is computed from the mean Poynting flux through the surfaces at $y = \pm 0.25 \, R$ ($x = \pm 0.9 \, R$, respectively), multiplied by their area. The kinetic power is computed from particles in a thin slab of thickness $w_s = 10 \, \Delta x$ adjacent to each surface, $P_\text{KE} = m_e c^3 / w_s \sum_j \left( \gamma_j - 1 \right) \beta^i_j$, where $\beta^i_j$ is the dimensionless velocity component of particle $j$ perpendicular to the surface, and $\gamma_j$ is the particle Lorentz factor. Radiative losses are computed from the rate of work done by the radiative drag forces on the $N$ particles within the volume, $L = - \sum_j^{N} \bm{F}_j \cdot \bm{\beta}_j c$, where $\bm{F}_j$ is the drag force acting on particle $j$. We further separate the outgoing kinetic power into contributions from particles with Lorentz factors below and above the SSA threshold, denoted by $P_\text{KE}^<$ and $P_\text{KE}^>$, respectively. The former includes the contribution of bulk motions.

\begin{figure*}[!t]
    \centering
    \includegraphics[width=0.9\textwidth]{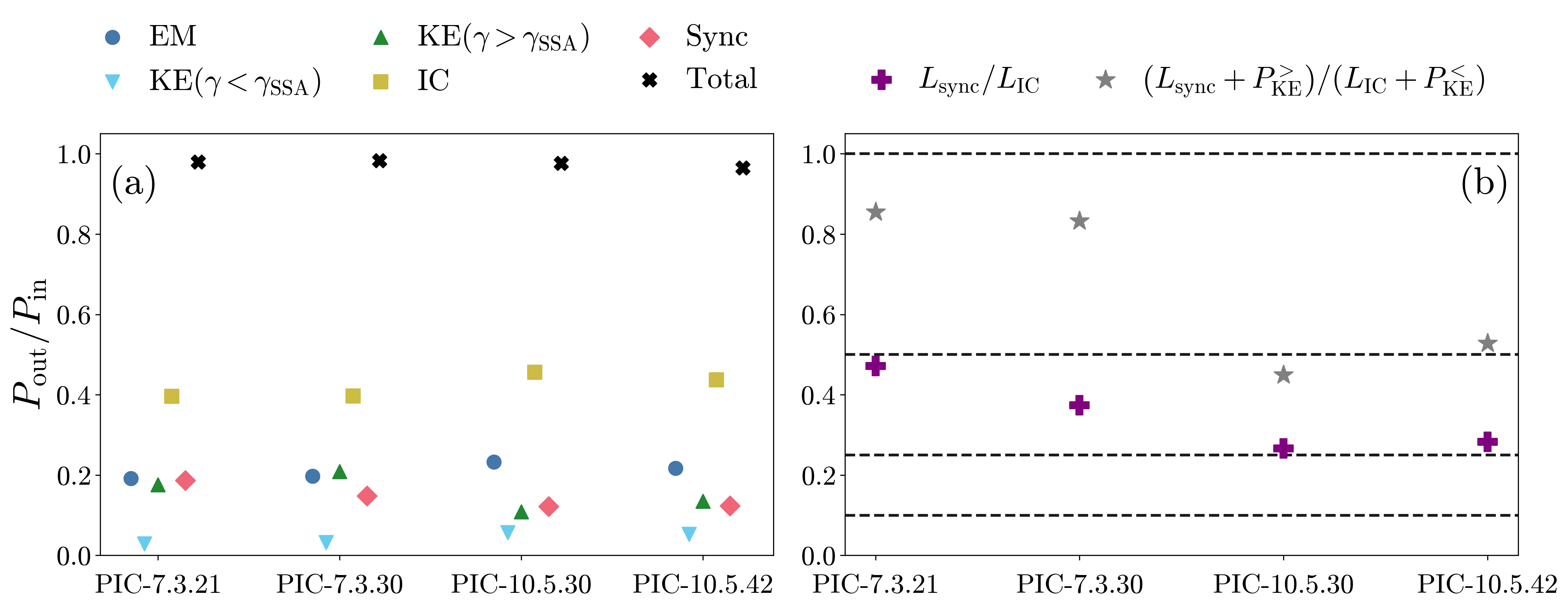}
    \caption{Panel (a) shows the time-averaged components of the outgoing power for all simulations that include both cooling channels. Data points for each simulation are shifted horizontally for clarity.   
    Panel (b) shows the ratio of synchrotron to IC power, which ranges from $0.25$ to $0.5$. It also shows an upper estimate of this ratio, accounting for the kinetic energy of escaping particles that could be radiated through the respective cooling channels (see text). These results motivate the ratios $0.1$, $0.25$, $0.5$, and $1$ (horizontal dashed lines) adopted in Section~\ref{sec:lehamoc}.\label{fig:power-avg}}
\end{figure*}

Figure~\ref{fig:power-time} shows the evolution of these quantities in \hybrid{}, normalized by the time-averaged total incoming power, $\langle P_\text{in} \rangle$. Since the kinetic energy flux of the upstream plasma is much smaller than the Poynting flux, the incoming power is $P_\text{in} \sim 2 c \beta_{\text{rec}} B^2 R_x R_z / (4 \pi)$, where the factor of two accounts for the two inflow sides of the reconnection layer. The nearly constant ratio $P_\text{in}/\langle P_\text{in} \rangle$ in Figure~\ref{fig:power-time} shows that the reconnection layer reaches a statistical steady state between $t_i = 3 \, R/c$ and $t_f = 10 \, R/c$. The normalized outgoing power, $P_\text{out}/\langle P_\text{in} \rangle$, also fluctuates around unity, indicating that the total energy leaving the reconnection region balances the energy entering it and further confirming the steady state. The largest fluctuations occur in $P_\text{EM}$, partly due to large magnetized flux ropes crossing the surfaces where the outgoing fluxes are measured.

Figure~\ref{fig:power-avg}(a) shows the time-averaged outgoing power in all the simulations that include both cooling channels. For simplicity, hereafter we omit the angle brackets when referring to time-averaged quantities. In all cases, approximately $20\%$ of $P_\text{in}$ leaves the system as Poynting flux, with a similar fraction carried away as particle kinetic energy. The remaining energy is radiated, with approximately $40\%$ of $P_\text{in}$ emitted through IC scattering and $20\%$ through synchrotron emission. The radiative efficiency of reconnection is therefore at least $\eta \simeq 0.6$. Including the residual particle kinetic energy, which would be radiated away in the fast-cooling limit (not fully realized in our simulations), the radiative efficiency could reach $\eta \simeq 0.8$.

Figure~\ref{fig:power-avg}(a) shows that $L_\text{IC}$ is slightly higher in \hybrid{} and \hybridsync{} than in  \hybridicstrong{} and \hybridic{}, despite their larger \gammaic{} and hence weaker IC cooling. The higher SSA threshold in the first two simulations allows more particles cooling via synchrotron emission to enter the IC-dominated regime while retaining some of their energy, which can then be radiated through IC scattering. The same reasoning explains why the synchrotron luminosity is higher in \hybridicstrong{} and \hybridic{}.

Particles with $\gamma < \gamma_\text{SSA}$ are in the fast-cooling regime: as shown in Figure~\ref{fig:power-avg}(a), their radiative losses exceed the outgoing kinetic power. For particles above the SSA threshold, the synchrotron power is comparable to the outgoing kinetic power. As discussed earlier, our simulations are only marginally in the fast-cooling regime for synchrotron emission, with $\tau_\text{cool}^\text{sync} \sim \gamma_\text{SSA}$. Reaching a deeper fast-cooling regime would require a larger simulation domain (see Section~\ref{sec:pic-cool} and Appendix~\ref{app:2k}). We therefore regard the computed synchrotron power as a lower limit. To estimate the synchrotron-to-IC power ratio $\xi$ in the fast-cooling limit, we assume that the residual kinetic energy carried by escaping particles is eventually radiated through the cooling channel that dominates at their escape energy. Specifically, we assign the outgoing kinetic power of particles with $\gamma > \gamma_\text{SSA}$, $P_\text{KE}^>$, to synchrotron emission, and that of particles with $\gamma < \gamma_\text{SSA}$, $P_\text{KE}^<$, to IC emission. This gives $\xi \simeq \left(L_\text{sync} + P_\text{KE}^> \right) / \left(L_\text{IC} + P_\text{KE}^< \right)$.

Figure~\ref{fig:power-avg}(b) shows that synchrotron emission accounts for a substantial fraction of the total radiative losses. Across our simulations, the lower limit on $\xi$ ranges from $0.25$ to $0.5$, while the upper estimate ranges from $0.5$ to $0.9$. In realistic BH coronae, where $\gamma_\text{SSA} \ll \sigma$, an even larger fraction of the dissipated energy could be radiated through synchrotron emission. As discussed in Section~\ref{sec:analytical}, the high-energy synchrotron photons undergo pair production, which modifies the escaping photon spectrum. We model this effect in the next section.

\section{Coronal photon spectrum}\label{sec:lehamoc}
We now follow how the photon spectrum injected by reconnection-accelerated particles (Section~\ref{sec:rad-rec-emission}) is reprocessed in the corona, for synchrotron-to-IC power ratios $\xi$ motivated by the PIC simulations (Section~\ref{sec:pic}). We compare the results with the analytical predictions developed in Section~\ref{sec:analytical}.

\subsection{\texttt{LeHaMoC} setup}
\begin{deluxetable}{lclc}
\tablecaption{Parameters of the injected spectrum that are fixed in all \texttt{LeHaMoC} models (Eqs.~(\ref{eq:L-tot-inj}), (\ref{eq:inj-ic}), and (\ref{eq:inj-sync})). Slopes refer to $\varepsilon L_\varepsilon \propto \varepsilon^{\text{slope}}$; model-dependent parameters ($\xi$, $R$, $L_2$) are listed in Table~\ref{tab:moc-xi}. All models use a coronal magnetic field $B = 10^{8}$~G; the injected spectrum spans $0.1 \, \text{keV}$--$100 \, \text{MeV}$.  \label{tab:moc-params}}
\tablehead{\multicolumn{2}{c}{IC component} & \multicolumn{2}{c}{Synchrotron component}}
\startdata
$L_1$           & $1.285\times10^{37}$ erg s$^{-1}$ & $\varepsilon_\text{br}$ & 100 keV \\
$\varepsilon_1$ & 159 keV                           & $\varepsilon_2$         & 45 MeV  \\
$a$             & 0.7                               & $\Gamma_<$              & 0.5     \\
                &                                   & $\Gamma_>$              & 0       \\
                &                                   & $k$                     & 0.2     \\
\enddata
\end{deluxetable}

We use the leptonic module of the lepto-hadronic modeling code \texttt{LeHaMoC} \citep{stathopoulos_lehamoc_2024}. We model the corona as a homogeneous sphere of radius $R$ and magnetic field strength $B = 10^8 \, \text{G}$ and continuously inject the photon spectrum of reconnection-accelerated particles, as described in Section~\ref{sec:rad-rec-emission}. The code accounts for $\gamma\gamma$ pair production, synchrotron emission, synchrotron self-absorption, and inverse Compton scattering by the secondary pairs. We evolve each model until both the photon spectrum and the secondary pair distribution reach a steady state, which happens $t \simeq 3 \, R/c$ after the start of the calculation.

We keep the parameters of the injected IC component fixed and consider $\xi = 0.1$, $0.25$, $0.5$, and $1$. For each value of $\xi$, we choose $R$ such that the total radiated luminosity equals a fraction $\eta$ of the magnetic energy $U_B V$ dissipated over the reconnection timescale $t_\text{rec} \sim R / (\beta_\text{rec}c)$. Motivated by our PIC simulations (Section~\ref{sec:pic-results}), we take $\eta = 0.8$, which gives
\begin{equation}\label{eq:lehamoc-R}
    R = \sqrt{\frac{6 L_\text{IC} (1 + \xi)}{\eta \beta_\text{rec} c B^2}} \simeq 2 \times 10^6 \sqrt{1 + \xi} ~\text{cm}.
\end{equation}
For all values of $\xi$, the resulting $R$ is consistent with the value of $R \sim 2 \, r_g$ adopted in Section~\ref{sec:rad-rec}. Table~\ref{tab:moc-xi} summarizes the parameters we vary, while the remaining parameters, which are held fixed, are listed in Table~\ref{tab:moc-params}. In Appendix~\ref{app:lehamoc-sync}, we test the sensitivity of our results to the parameters of the injected synchrotron component.

\begin{table}
\centering
\begin{tabular}{cccc}
\hline
Model & $\xi$ & $L_2 \, \left[ \text{erg/s}\right]$ & $R$ [$10^6\,$cm] \\
\hline
\sedlo{} & 0.1 & $2.2 \times 10^{35}$ & $2.10$ \\
\sedmid{} & 0.25 & $5.58 \times 10^{35}$ & $2.25$ \\
\sedfiducial{} & 0.5 & $1.12 \times 10^{36}$ & $2.5$ \\ 
\sedhi{} & 1 & $2.23 \times 10^{36}$ & $2.85$ \\
\hline
\end{tabular}
\caption{\texttt{LeHaMoC} models with different ratios $\xi$ of synchrotron to IC power; \sedfiducial{} is the fiducial model. $L_2$ is the normalization of the injected synchrotron component, and $R$ is the size of the system, see Eq.~(\ref{eq:lehamoc-R}). All other parameters are as in Table~\ref{tab:moc-params}.}
\label{tab:moc-xi}
\end{table}

\subsection{\texttt{LeHaMoC} results}
\subsubsection{Fiducial model}
\begin{figure}
    \centering
    \includegraphics[width=\columnwidth]{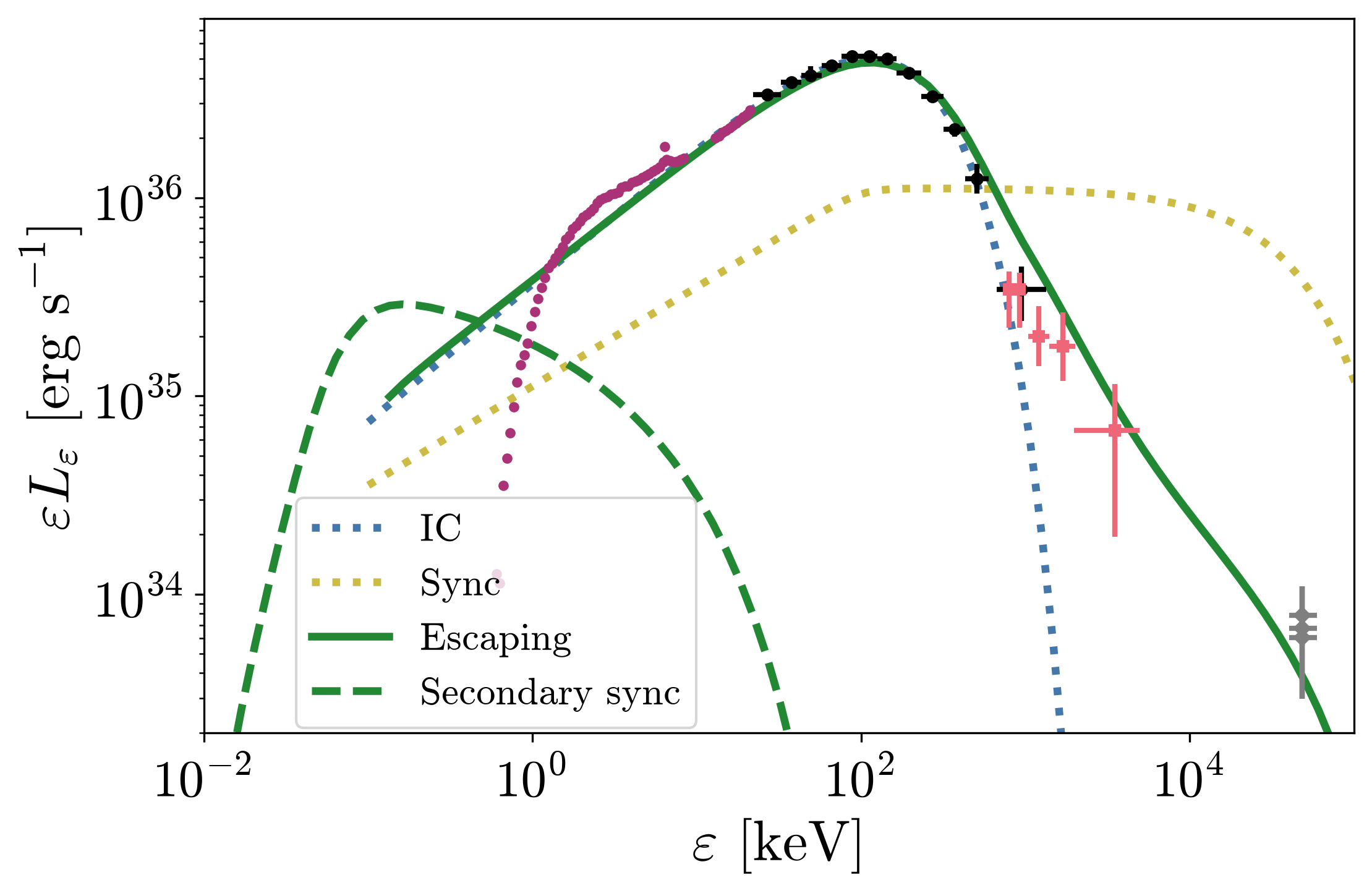}
    \caption{Steady-state photon spectrum of the fiducial model \sedfiducial{}. The solid line corresponds to the escaping injected spectrum (attenuated at high energies), while the dashed line shows the synchrotron spectrum of cooling secondary pairs. Dotted lines show the injected IC and synchrotron components; observations of the hard state of Cygnus~X-1 are as in Figure~\ref{fig:inj-spec}.}
    \label{fig:sed-fiducial}
\end{figure}

Figure~\ref{fig:sed-fiducial} shows the steady-state photon spectrum of the fiducial model \sedfiducial{}. Pair production suppresses the injected synchrotron component at $\varepsilon \gtrsim \varepsilon_\text{tail} \simeq 1.5 \, \text{MeV}$, consistent with our estimate in Section~\ref{sec:analytical-tau} that the corona becomes optically thick to $\gamma\gamma$ absorption above this energy. As expected from Section~\ref{sec:analytical-tail}, $\gamma\gamma$ absorption steepens the spectrum relative to the injected synchrotron component: above $\varepsilon_\text{tail}$, the attenuated spectrum follows a power law, $\varepsilon L_\varepsilon \propto \varepsilon^{-1.2}$. The overall shape of the photon spectrum above $\sim 1 \, \text{MeV}$ is consistent with the MeV emission observed from Cygnus~X-1 in the hard state \citep{McConnell2000,2017MNRAS.471.3657Z}.

The secondary pairs cool via synchrotron emission, producing an additional component at low energies. This component becomes optically thin above $\varepsilon_{\rm SSA} \simeq 0.13 \, \text{keV}$, where its spectrum peaks (Figure~\ref{fig:sed-fiducial}). The SSA photon energy corresponds to a particle Lorentz factor $\gamma_\text{SSA} \sim 10$. These values are in reasonable agreement with the analytical estimates in Section~\ref{sec:analytical-pm-spec}, where we found $\varepsilon_\text{SSA} \simeq 0.035 \, \text{keV}$ and $\gamma_\text{SSA} \simeq 5$.

The secondary synchrotron photons can serve as seed photons for Comptonization in the reconnection layer. As shown in Figure~\ref{fig:sed-fiducial}, at the SSA energy $\varepsilon_{\rm SSA}$ the secondary synchrotron component exceeds the modeled IC spectrum by a factor of $\sim 3$. This is consistent with the prediction of Equation~(\ref{eq:xi-seed-photons}), since $\xi = 0.5 > \xi_s$. Our results therefore suggest that secondary synchrotron emission can supply the seed photons for Comptonization in BH coronae.

\subsubsection{Dependence on $\xi$}
\begin{figure}
    \centering
    \includegraphics[width=\columnwidth]{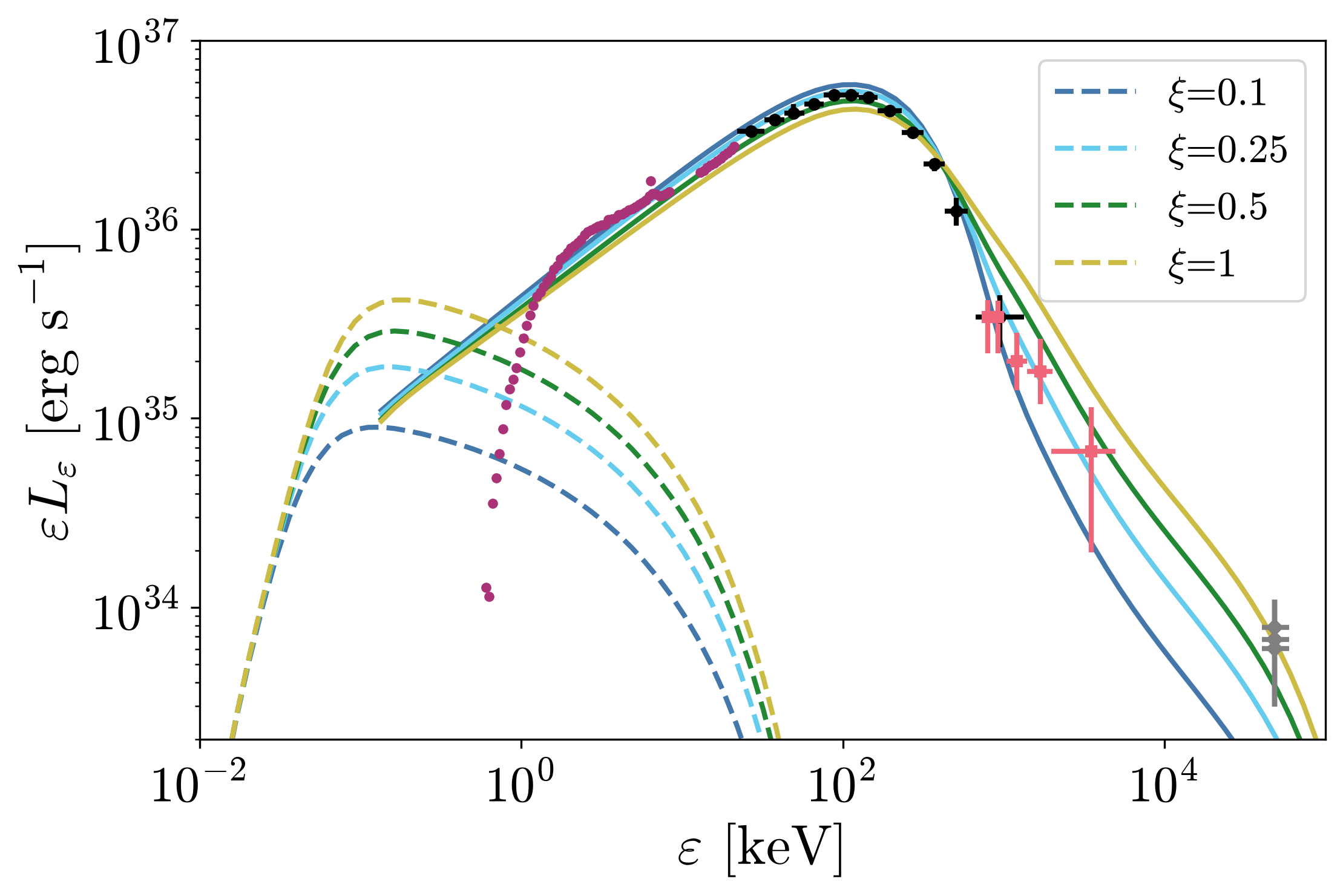}
    \caption{Steady-state photon spectra for $\xi=0.1$, $0.25$, $0.5$, and $1$. As in Figure~\ref{fig:sed-fiducial}, solid and dashed lines correspond to the attenuated injected spectrum and the secondary synchrotron emission, respectively. The escaping spectrum extends to the synchrotron burnoff limit.}
    \label{fig:sed-xi}
\end{figure}

Figure~\ref{fig:sed-xi} shows the steady-state photon spectra for different values of $\xi$, which controls the normalization of the injected synchrotron component (Table~\ref{tab:moc-xi}). We first discuss the MeV tail and then the synchrotron emission from cooling secondary pairs.

The slope of the MeV tail is nearly independent of $\xi$, since it depends only on the shape of the injected spectra (Section~\ref{sec:analytical-tail}). By contrast, the luminosity of the tail depends on $\xi$, with its normalization scaling as $\varepsilon L_\varepsilon^\gamma \propto \xi R \propto \xi \sqrt{1 + \xi}$ (Eqs.~(\ref{eq:Lg-ss}) and (\ref{eq:lehamoc-R})). This scaling predicts an increase by a factor of $\simeq 14$ from \sedlo{} to \sedhi{}, compared to the factor of $\simeq 9$ obtained numerically, in reasonable agreement. The observed MeV tail is best reproduced for $\xi = 0.25$--$0.5$, consistent with the range of $\xi$ measured in our PIC simulations.

Turning to synchrotron emission from secondary pairs, we expect that higher $\xi$ results in more energy being deposited into the secondary pairs (approximately $10$ times more in \sedhi{} than in \sedlo{}). Consequently, the secondary synchrotron component becomes more prominent, with its peak luminosity increasing by a factor of $\sim 5$ from \sedlo{} to \sedhi{}. The peak energy, $\varepsilon_\text{SSA} \simeq 0.13 \, \text{keV}$, depends only weakly on $\xi$, consistent with the scaling $\varepsilon_\text{SSA} \propto \xi^{2/7}$ in Equation~(\ref{eq:eps-ssa}).

Figure~\ref{fig:sed-xi} shows that the secondary synchrotron peak in \sedlo{} lies below the injected IC component, whereas at higher $\xi$, secondary synchrotron photons could provide the dominant seed photon supply for Comptonization. We therefore infer a numerical threshold of $\xi_s^\text{num} \simeq 0.1$, approximately five times the analytical estimate $\xi_s$ derived in Section~\ref{sec:analytical-pm-spec}. This discrepancy arises because the analytical calculation overestimates the peak synchrotron luminosity by a factor of $\sim 3$--$5$, possibly due to the approximate $\gamma\gamma$ cross-section adopted there, which overestimates the true cross-section near the pair-production threshold.

\section{Discussion}\label{sec:disc}
\subsection{Implications for polarization}
The polarization degree (PD) of the hard-state emission of Cygnus~X-1 depends on photon energy. It is $\sim 4\%$ at $2$--$8$~keV \citep{kraw_22} and $\sim 5\%$ at $19$--$64$~keV \citep{2025ApJ...994...37A}. Above a few hundred keV, the PD has been reported to reach $\sim 70$--$75\%$ \citep{2011Sci...332..438L,2012ApJ...761...27J,2015ApJ...807...17R}, although further observations, for example with COSI, are needed to establish these measurements more firmly \citep{2025ApJ...994...37A}.

Such a high PD in the MeV tail would be consistent with synchrotron emission from non-thermal particles \citep{2026arXiv260806328B}. Indeed, it has often been suggested that the MeV emission originates in the jet rather than the corona (e.g., \citealt{2014MNRAS.442.3243Z}).

We demonstrate that a single physical mechanism---relativistic magnetic reconnection---operating within the corona can account for both the $\sim 100$~keV peak and the MeV tail. The former arises from bulk Comptonization, as established in previous studies \citep[e.g.,][]{sridhar_comptonization_2021}, while the latter is produced by synchrotron emission from particles rapidly accelerated through alternating free and trapped phases \citep[e.g.,][]{zhang_fast_2021}. Our results show that synchrotron $\gamma$-rays escaping the corona can naturally explain the MeV tail, without requiring a separate emission region in the jet. Given its synchrotron origin, this component could also exhibit a high degree of polarization. Quantifying the PD of the escaping, $\gamma\gamma$-attenuated synchrotron tail is an important goal for future work.

Our model also has implications for the polarization properties of the Comptonized component. \citet{2026ApJ..1008L..37G} recently reproduced the IXPE polarization measurements of Cygnus~X-1 \citep{kraw_22} by injecting seed photons near the mid-plane of the reconnection layer. Such a localized injection of seed photons arises naturally in our scenario, where synchrotron emission from cooling secondary pairs provides the seeds for Comptonization, whereas it is less natural if the seed photons originate in the accretion disk, as is often invoked. Future work should also investigate whether, and to what extent, the polarization properties of the seed photons are preserved during Comptonization.

\subsection{Limitations}
Our model assumes a highly magnetized ($\sigma = 10$) pair plasma. However, the magnetic field strength and plasma composition in BH coronae remain poorly constrained (see Section~\ref{sec:rad-rec}). At lower magnetizations, $\sigma \lesssim 1$, less magnetic energy per particle is available for dissipation, resulting in lower thermal and bulk energies. This could make it difficult to reproduce the $\sim 100$~keV peak, challenging reconnection-based models in this regime \citep{sridhar_comptonization_2023}. At low magnetizations, the corona may also contain a significant proton component, making the partition of dissipated energy among particle species an important additional ingredient (e.g., \citealt{hakobyan_reconnection_2025,2026arXiv260921118M}). Future studies should investigate the dependence of our results on both magnetization and plasma composition.

We have also assumed a very weak guide field, whereas relativistic reconnection may occur in the strong-guide-field regime, for example within turbulence. A strong guide field suppresses the ``free'' acceleration of particles \citep{2026arXiv261003427K}, thus reducing the number of particles that emit high-energy synchrotron photons. An extension of our study to the strong-guide-field regime will be presented elsewhere.

Due to the computational cost of 3D PIC simulations, we had to reduce the separation of scales, as compared to realistic BH coronae. As discussed in Section~\ref{sec:pic}, synchrotron-emitting particles above the SSA threshold are not fully in the fast-cooling regime, so the synchrotron power measured in our simulations is likely a lower limit. Future work will investigate how the partition of dissipated energy among the different cooling channels scales, as we approach the limit of asymptotically large scale separation (see also Appendix~\ref{app:2k}).

Two major simplifications underlie our analytical treatment in Section~\ref{sec:analytical} and the radiative calculations of Section~\ref{sec:lehamoc}. First, we assume a spherical coronal geometry, whereas the reconnection layer more closely resembles a slab. Second, the anisotropy of the photon populations---neglected in our models---may affect the pair-production rate and the properties of the secondary pairs, including their pitch angle distribution. These, in turn, determine how efficiently the secondary pairs cool via synchrotron emission and whether they produce enough seed photons for Comptonization. Capturing these effects self-consistently would require PIC simulations that include not only IC and synchrotron cooling but also pair production.

\section{Summary}\label{sec:sum}
We have revisited the reconnection-based model of the hard state of X-ray binaries \citep{beloborodov_radiative_2017,sironi_kinetic_2020,sridhar_comptonization_2021}, using Cygnus~X-1 as a representative example. We incorporate recent advances in our understanding of 3D reconnection physics, which show that some particles can undergo rapid acceleration  and reach much higher energies than in 2D \citep{zhang_fast_2021,zhang_origin_2023,Bowyer2026}. As a result, a significant fraction of the dissipated magnetic energy can be radiated as synchrotron emission, with a spectrum  extending up to the synchrotron burnoff limit of $\sim 16$~MeV, and potentially beyond. We model this radiation and show that the highest-energy synchrotron photons undergo pair production, generating secondary pairs that cool via synchrotron emission and supply seed photons for Comptonization. Our main results are as follows:
\begin{itemize}
    \item Synchrotron emission is a significant cooling channel in BH coronae. In our PIC simulations, $\sim 40\%$ of the incoming electromagnetic power is radiated via IC scattering and $\sim 10$--$20\%$ via synchrotron emission. The resulting synchrotron-to-IC power ratio is $\xi \simeq 0.25$--$0.5$. This is a lower limit, since particles above the SSA threshold are only marginally in the fast-cooling regime in our simulations. If all particles radiated their remaining energy before escaping, as expected for coronal conditions, the ratio could reach $\xi \simeq 0.5$--$0.9$.
    \item The MeV tail originates from synchrotron emission by reconnection-accelerated particles, attenuated by pair production. Our analytical estimates and numerical calculations show that the corona is optically thick to $\gamma\gamma$ absorption above $\varepsilon \sim 1.5 \, \text{MeV}$. The escaping synchrotron photons form an MeV tail consistent with observations of Cygnus~X-1 up to tens of MeV. The observed emission is best reproduced for $\xi = 0.25$--$0.5$.
    \item We analytically estimate that secondary pairs contribute a Thomson optical depth of order unity, suggesting that they can account for most, if not all, of the pairs in the reconnection region.
    \item Synchrotron emission from secondary pairs can provide the seed photons for Comptonization. The secondary synchrotron spectrum peaks at $\varepsilon_\text{SSA} \simeq 0.13 \, \text{keV}$, with only a weak dependence on $\xi$. For $\xi > 0.1$, this component can potentially supply all the seed photons required for Comptonization.
\end{itemize}

Our results extend the reconnection-based model of the hard-state emission from X-ray binaries. Previous work has shown that bulk Comptonization in the reconnection layer can explain the $\sim 100 \, \text{keV}$ peak. We demonstrate that the same physical mechanism---magnetic reconnection---can also produce the MeV tail and supply the seed photons for Comptonization. In our model, both the seed photons and the escaping $\gamma$-rays originate from synchrotron emission and could therefore be highly polarized. Modeling their polarization properties could provide an important observational test of our scenario.

\section*{Acknowledgments}
We thank Andrei Beloborodov for insightful conversations. V.R. acknowledges the use of Claude (Claude Sonnet 5 and Opus 5.5, Anthropic) to refine the visualization scripts and suggest language improvements to the manuscript. Responsibility for the final manuscript lies entirely with the authors. L.S. acknowledges support from DoE Early Career Award DE-SC0023015, NASA ATP 80NSSC24K1238, NASA ATP 80NSSC24K1826, and NSF AST-2307202. This research was supported by the Multimessenger Plasma Physics Center (MPPC, NSF grant PHY-2206609) and by a grant from the Simons Foundation (MP-SCMPS-0000147).

\appendix
\section{Dependence on \taucoolic{} and \gammassa{} in PIC simulations}\label{app:2k}
\begin{figure}[!ht]
    \centering
    \includegraphics[width=0.9\textwidth]{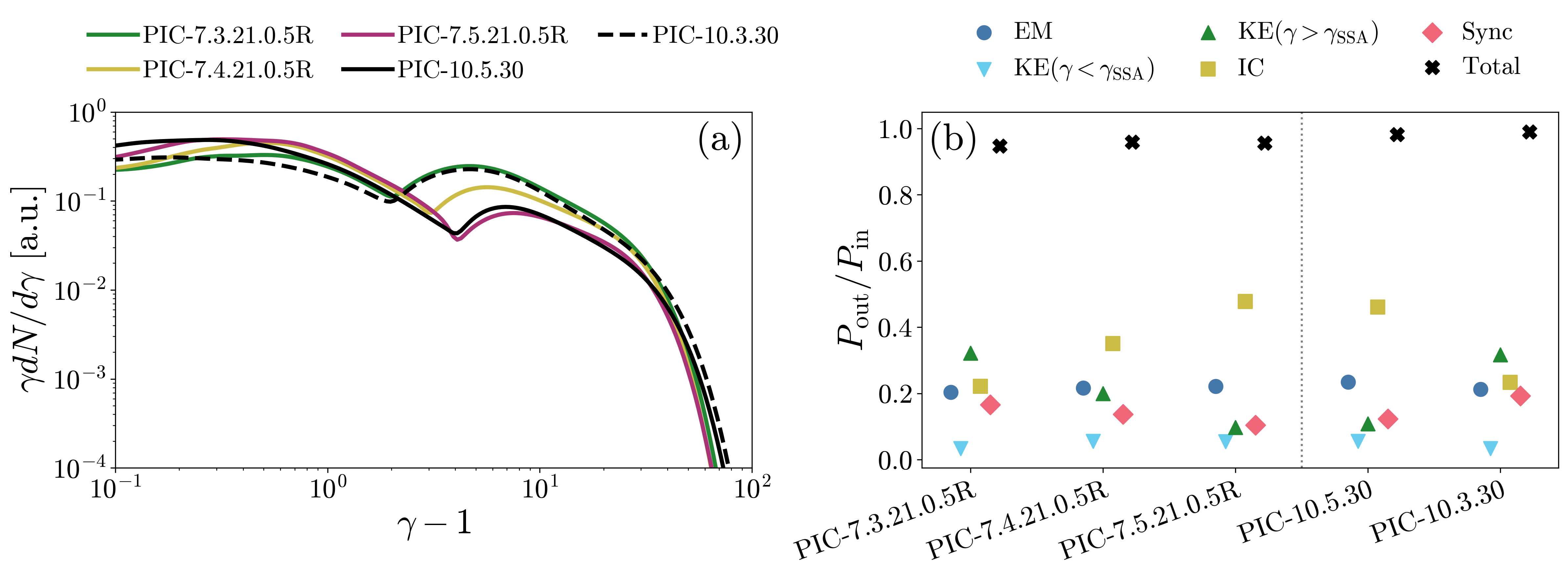}
    \caption{Comparison of the standard-box runs \hybrid{} and \hybridssa{} with three smaller-box simulations, labeled \texttt{0.5R}. Panels show the time-averaged (a) particle spectra and (b) power partitions. The \texttt{0.5R} simulations have the same \taucool{} as the larger-box simulations (for both synchrotron and IC cooling) and are time-averaged over $2 \le ct/R \le 3.25$, corresponding to $4$--$6.5$ light-crossing times of the smaller box. \label{fig:2k}}
\end{figure}
We use smaller-box PIC simulations to investigate the role of \taucoolic{} and \gammassa{}. According to Equation~(\ref{eq:scales-tau-cool}), \taucoolic{} depends on \gammaic{} and \gammamax{}, the latter scaling linearly with the system size $R$. We compare the standard-box simulations \hybrid{} and \hybridssa{} with three simulations in boxes of half the size ($0.5 \, R$): \twokfive{}, \twokfour{}, and \twokthree{}. In these smaller-box simulations, we decrease \gammarad{} by a factor of $\sqrt{2}$ to keep \taucool{} unchanged for both synchrotron and IC cooling.

Figure~\ref{fig:2k} shows that \hybrid{} and \twokfive{} have nearly identical particle spectra and energy partitions, as do \hybridssa{} and \twokthree{}. This suggests that our results depend on the strength of IC cooling only through \taucoolic{}. More specifically, at fixed $\sigma$ and $\gamma_{\rm SSA}$, the energy partition among different channels appears to be determined uniquely by  $\tau_\text{cool}^\text{IC}$ and $\tau_\text{cool}^\text{sync}$.

As discussed in Section~\ref{sec:pic-cool}, we expect \gammaic{} $\gg \sigma$ in BH coronae, but adopt \gammaic{} $\leq \sigma$ due to computational constraints. The agreement between \hybrid{}, with $\gamma_\text{rad}^\text{IC} = \sigma$, and \twokfive{}, with $\gamma_\text{rad}^\text{IC} < \sigma$, demonstrates that this choice does not affect our conclusions. In both simulations, particles reach \gammassa{} before \gammaic{}, and synchrotron emission becomes the dominant cooling channel above \gammassa{}. Thus, whether \gammaic{} lies above or below $\sigma$ has little effect on our results, provided that \gammassa{} $<$ \gammaic{} and particles below the SSA threshold are in the fast-cooling regime.

At fixed \taucoolic{} and \taucoolsync{}, the value of \gammassa{} determines how many particles fall within the IC-cooling range and hence the IC power. Figure~\ref{fig:2k}(b) shows that $L_{\rm IC}$ increases by a factor of $2$ from \twokthree{} to \twokfive{}, with \twokfour{} yielding an intermediate value (and the synchrotron power decreasing accordingly). We choose $\gamma_\text{SSA} = 3, \, 5$ to keep \gammassa{} below both \gammaic{} and $\sigma$, while maintaining a sufficiently broad range of Lorentz factors in the IC-cooling regime (i.e., between unity and $\gamma_{\rm SSA}$). In realistic BH coronae, however, both \gammaic{} and $\sigma$ are expected to greatly exceed \gammassa{}. Since the fraction of dissipated energy radiated through IC scattering is sensitive to \gammassa{}, future PIC simulations with a more realistic separation between \gammassa{},  $\sigma$, and \gammaic{} are needed to better constrain the partition of radiated power between IC and synchrotron emission.

\section{Dependence on the injected synchrotron spectrum in \texttt{LeHaMoC}}\label{app:lehamoc-sync}
We investigate how the emerging spectrum in Section~\ref{sec:lehamoc} depends on two parameters of the injected synchrotron component: the break energy $\varepsilon_{\rm br}$ and the cutoff energy $\varepsilon_2$ (Eq.~(\ref{eq:inj-sync})). We vary one parameter at a time, adjusting the normalization $L_2$ to keep the total synchrotron luminosity fixed. All other parameters are the same as in \sedfiducial{} (Tables~\ref{tab:moc-params} and \ref{tab:moc-xi}). The models are listed in Table~\ref{tab:sync-params}.

\begin{table}
\centering
\begin{tabular}{cc}
\hline
Varied parameter & $L_2 \, \left[ \text{erg/s}\right]$ \\
\hline
$\varepsilon_\text{br} = 10 \,$keV &  $8.88 \times 10^{35}$\\
$\varepsilon_\text{br} = 300 \,$keV & $1.30 \times 10^{36}$ \\
$\varepsilon_2 = 15 \,$MeV & $1.31 \times 10^{36}$\\
$\varepsilon_2 = 90 \,$MeV & $1.04 \times 10^{36}$\\
\hline
\end{tabular}
\caption{Models used to test the dependence of the resulting spectrum on the injected synchrotron component. Parameters not listed are as in \sedfiducial{} (Tables~\ref{tab:moc-params} and~\ref{tab:moc-xi} in the main text).}
\label{tab:sync-params}
\end{table}

\begin{figure*}
    \centering
    \includegraphics[width=0.9\textwidth]{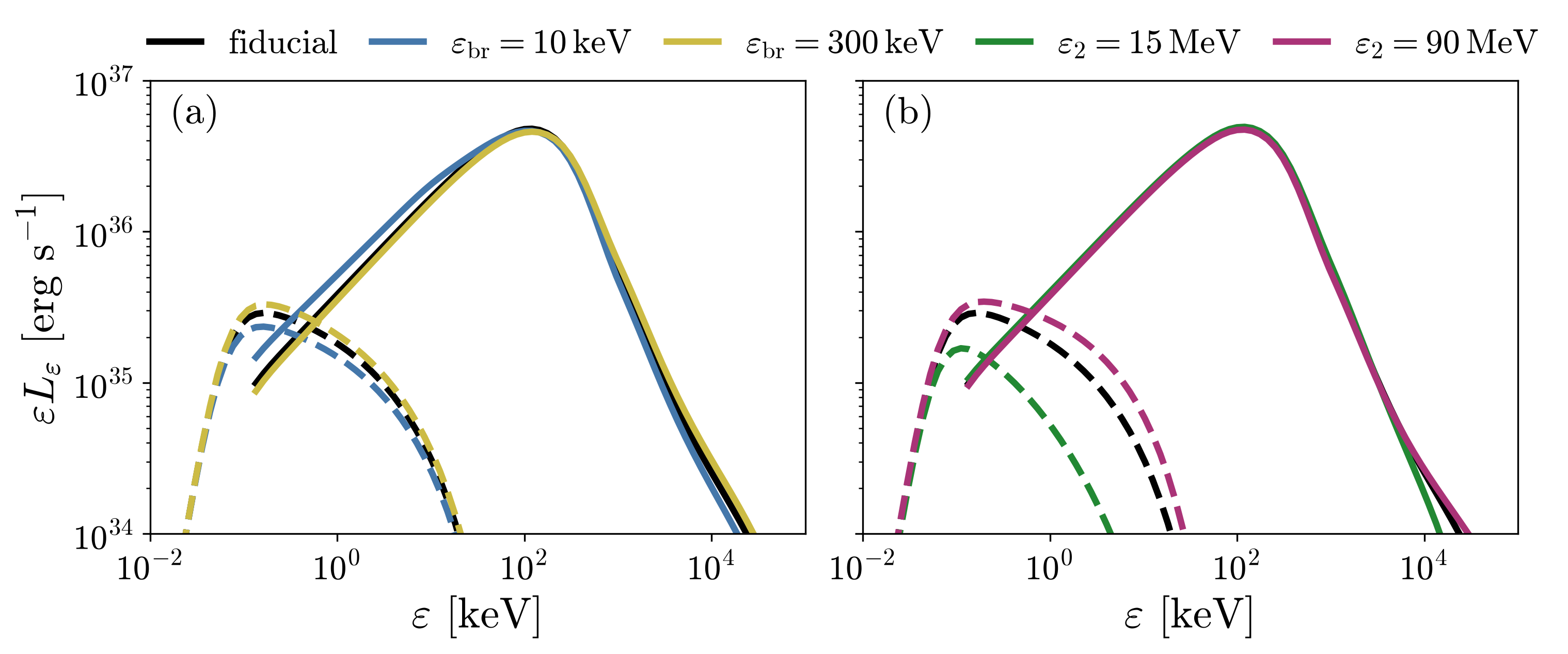}
    \caption{Steady-state photon spectra of models in Table~\ref{tab:sync-params}. Panels show the effect of varying (a) the synchrotron break energy, $\varepsilon_{\rm br}$, and (b) the synchrotron cutoff energy, $\varepsilon_2$. Solid lines show the reprocessed injected spectrum and dashed lines show the synchrotron emission of secondary pairs.}
    \label{fig:sync-params}
\end{figure*}

Figure~\ref{fig:sync-params}(a) shows that varying the break energy between $10$ and $300$~keV has little effect on the emerging spectrum: both the escaping $\gamma$-ray and secondary synchrotron spectra are nearly identical across all three models. This is expected, since both the attenuated synchrotron spectrum and the properties of the secondary pairs are governed by $\gamma\gamma$ pair production, which becomes important only above $\sim 1.5$~MeV, well above $\varepsilon_{\rm br}$.

The classical radiation-reaction limit gives $\varepsilon_2 \sim 16$~MeV for $\beta_{\rm rec}=0.1$ (Section~\ref{sec:rad-rec-emission}), but variations in the reconnection rate by a factor of a few or pitch-angle effects could modify this value by a factor of order unity. We therefore compare the fiducial model \sedfiducial{} with models having $\varepsilon_2 = 15$~MeV and $\varepsilon_2 = 90$~MeV in Figure~\ref{fig:sync-params}(b). The escaping $\gamma$-ray spectra (solid lines) are nearly identical across all three models, up to tens of MeV. By contrast, $\varepsilon_2$ affects the amount of power transferred to the secondary synchrotron component (dashed lines). Nevertheless, in all cases, the secondary synchrotron emission is sufficiently luminous to supply seed photons for Comptonization. Thus, for the fiducial synchrotron-to-IC power ratio, $\xi = 0.5$, secondary synchrotron photons remain a viable source of seed photons across the explored range of $\varepsilon_2$.

\bibliography{references}{}
\bibliographystyle{aasjournalv7}
\end{document}